\documentclass[fleqn,usenatbib]{mnras}

\usepackage{newtxtext,newtxmath}
\usepackage[T1]{fontenc}
\usepackage{ae,aecompl}

\usepackage{graphicx}	% Including figure files
\usepackage{amsmath}	% Advanced maths commands
\usepackage{amssymb}	% Extra maths symbols
\usepackage{tikz}

\newcommand{\hmpc}{\,$h$\,Mpc$^{-1}$}
\newcommand{\mpch}{\,Mpc\,$h^{-1}$}
\newcommand{\arepo}{\textsc{Arepo}}
\newcommand{\music}{\textsc{MUSIC}}
\newcommand{\classcode}{\textsc{CLASS}}
\newcommand{\camb}{\textsc{CAMB}}

\title[ETHOS -- effective parametrization at high-z]{ETHOS -- An effective parametrization and classification for structure formation: the non-linear regime at $z\gtrsim5$}

\author[S. Bohr et al.]{Sebastian Bohr$^{1}$\thanks{E-mail: \href{mailto:seb21@hi.is}{seb21@hi.is}}, Jes{\'u}s Zavala$^{1}$, Francis-Yan Cyr-Racine$^{2}$,\newauthor Mark Vogelsberger$^{3}$, Torsten Bringmann$^{4}$ and Christoph Pfrommer$^{5}$
\\
$^{1}$Centre for Astrophysics and Cosmology, Science Institute, University of Iceland, Dunhagi 5, 107 Reykjavik, Iceland \\
$^{2}$Department of Physics and Astronomy, University of New Mexico, Albuquerque, NM 87131, USA \\
$^{3}$Department of Physics, Kavli Institute for Astrophysics and Space Research, Massachusetts Institute of Technology, Cambridge, MA 02139, USA \\
$^{4}$Department of Physics, University of Oslo, Box 1048, N-0371 Oslo, Norway \\
$^{5}$Leibniz-Institut f{\"u}r Astrophysik Potsdam, An der Sternwarte 16, 14482 Potsdam, Germany 
}

\date{Accepted XXX. Received YYY; in original form ZZZ}

\pubyear{2020 }

\begin{document}
\label{firstpage}
\pagerange{\pageref{firstpage}--\pageref{lastpage}}
\maketitle

\begin{abstract}
We propose two effective parameters that fully characterise galactic-scale structure formation at high redshifts ($z\gtrsim5$) for a variety of dark matter (DM) models that have a primordial cutoff in the matter power spectrum. Our description is within the recently proposed ETHOS framework and includes standard thermal Warm DM (WDM) and models with dark acoustic oscillations (DAOs). To define and explore this parameter space, we use high-redshift zoom-in simulations that cover a wide range of non-linear scales from those where DM should behave as CDM ($k\sim10$\hmpc), down to those characterised by the onset of galaxy formation ($k\sim500$\hmpc). We show that the two physically motivated parameters $h_{\rm peak}$ and $k_{\rm peak}$, the amplitude and scale of the first DAO peak, respectively, are sufficient to parametrize the linear matter power spectrum and classify the DM models as belonging to effective non-linear structure formation regions. These are defined by their relative departure from Cold DM ($k_{\rm peak}\rightarrow\infty$) and WDM ($h_{\rm peak}=0$) according to the non-linear matter power spectrum and halo mass function. We identify a region where the DAOs still leave a distinct signature from WDM down to $z=5$, while a large part of the DAO parameter space is shown to be degenerate with WDM. Our framework can then be used to seamlessly connect a broad class of particle DM models to their structure formation properties at high redshift without the need of additional $N$-body simulations.  
\end{abstract}


%%%%%%%%%%%%%%%%%%%%%%%%%%%%%%%%%%%%%%%%%%%%%%%%%%

%%%%%%%%%%%%%%%%% BODY OF PAPER %%%%%%%%%%%%%%%%%%

\section{Introduction}

Dark matter (DM) is a crucial ingredient in the formation of structures in the Universe as it makes up the majority of its matter content. Although the most likely explanation for DM is the particle hypothesis, its specific nature remains a mystery. The CDM model of structure formation has now emerged as the standard paradigm, and it has been shown to be consistent with the observed large scale structure of the Universe \citep[e.g.][]{Springel2005}. At smaller (galactic) scales however, the CDM model has faced a number of significant challenges over the last decades: (i) the underabundance of low-mass galaxies (either satellites or in the field) \citep{Klypin1999,Moore1999,Zavala2009,Papastergis2011,Klypin2015}, 
(ii) the core-cusp problem in low-surface brightness galaxies and possibly in dwarf spheroidals \citep{deBlok1997,Walker2011},
(iii) the "too-big-to-fail problem" \citep{Boylan-Kolchin2011,Papastergis2015}, (iv) the plane of satellites problem \citep{Pawlowski2013}, and (v) the diversity problem of rotation curves in dwarf galaxies \citep{Oman2015}. We note that with recent observations of ultra-faint galaxies, the too-big-to-fail problem becomes a diversity problem as well for the broad distribution of stellar kinematics in dwarf spheroidals in the Milky Way \citep{Zavala2019b}.

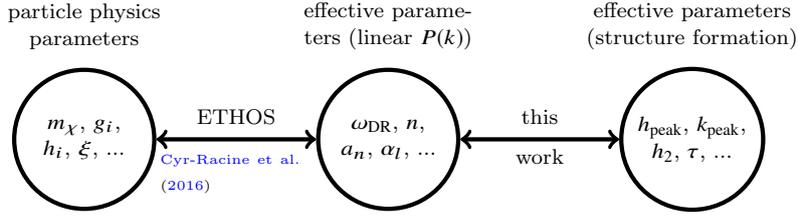
\begin{figure*}
    \centering
    \begin{tikzpicture}
        \node (particle) [ultra thick,circle,draw=black,minimum width=1.7cm,text centered,text width=1.5cm]{$m_\chi$, $g_i$, $h_i$, $\xi$, ...};
        \node[above of=particle,text centered,text width=3cm,yshift=.5cm] {particle physics parameters};
        \node (linear) [right of=particle,ultra thick,circle,draw=black,xshift=3cm,minimum width=1.7cm,text centered,text width=1.5cm] {$\omega_{\rm DR}$, $n$, $a_n$, $\alpha_l$, ...};
        \node[above of=linear,text centered,text width=3cm,yshift=.5cm] {effective parameters (linear $P(k)$)};
        \node (nonlinear) [right of=linear,ultra thick,circle,draw=black,xshift=3cm,minimum width=1.7cm,text centered,text width=1.5cm]{$h_{\rm peak}$, $k_{\rm peak}$, $h_2$, $\tau$, ...};
        \node[above of=nonlinear,text centered,text width=3cm,yshift=.5cm] {effective parameters (structure formation)};
        \draw [ultra thick,<->] (particle) -- node[anchor=south,yshift=2] {ETHOS} node[anchor=north,yshift=-2,text width=2cm] {\tiny \citet{Cyr-Racine2016}} (linear);
        \draw [ultra thick,<->] (linear) -- node[anchor=south,yshift=2] {this} node[anchor=north] {work} (nonlinear);
    \end{tikzpicture}
    \caption[]{Diagram illustrating the different sets of parameters that characterise a given DM particle model and the connections between them within the ETHOS framework, both in \citet{Cyr-Racine2016} and in this work. The particle physics space parameters such as the DM particle mass $m_\chi$, coupling constants $g_i$ (e.g. DM-DR), internal parameters $h_i$ such as the mediator mass and degrees of freedom and the present day DR to CDM temperature ratio $\xi$, were mapped in \cite{Cyr-Racine2016} into effective parameters fully describing the linear DM power spectrum (see section~\ref{sec:connection}). In this work, we make a re-parametrization, defining new ETHOS parameters that have both a more straightforward interpretation in terms of the linear power spectrum and a clearer physical interpretation (amplitude $h_{\rm peak}$ and scale $k_{\rm peak}$ of the first DAO peak, amplitude of the second peak $h_2$, and damping of higher order peaks $\tau$; see section~\ref{sec:parametrization}). The redefined ETHOS parameter space can be connected naturally to that defined in section \ref{sec:connection}, and thus to the particle physics space. Crucially, it is also sufficient to characterise non-linear structure formation for a variety of relevant DM models (such as WDM and models with DAOs) in the high-redshift Universe.}
    \label{fig:cartoon}
\end{figure*}

There is a long history of attempts to provide a satisfactory solution to these issues based on either: (i) incompleteness, biases and systematic uncertainties in observations \citep[e.g.][for the ``missing satellites problem"]{Koposov2008,Kim2018}, (ii) invoking a strong influence of uncertain baryonic physics in dwarf galaxies (e.g. impulsive supernova feedback to explain DM cores \citep{Pontzen2012}, tidal effects from the Milky-Way disk to alleviate the too-big-to-fail problem \citep{GK2019}, 
%\CP{suppression of the star formation through supernova, \citealt{Vogelsberger2014}, or cosmic ray feedback \citealt{Pfrommer2017}}, 
and suppression of galaxy formation at the dwarf mass scale due to cosmic reionisation \citep{Gnedin2000,Sawala2016b} to explain the underabundance of low-mass galaxies); and (iii) additional DM physics, i.e. departures from the CDM hypothesis such as: Warm Dark Matter (WDM; for a review see \citealt{Adhikari2017}) where the relativistic motion of the DM particles in the early Universe reduces the abundance and inner DM densities of galactic-scale haloes relative to CDM \citep[e.g.][]{Colin2000,Lovell2012}; self-interacting DM (SIDM; for a review see \citealt{Tulin2018}) where DM particles have strong self-interactions redistributing energy in the centre of haloes, thus resulting in DM cores \citep[e.g.][]{Spergel2000,Vogelsberger2012}; and quantum effects at galactic scales if DM is made of extremely light bosons with $\mathcal{O}(1\,{\rm kpc})$ de Broglie wavelength (fuzzy DM; for a review see \citealt{Hui2017}), also giving rise to extended DM cores \citep{Robles2012,Mocz2017}. 

Whether the CDM challenges are due to missing new DM physics, systematic uncertainties, or an inaccurate account of baryonic physics remains an open question (for a recent review on  different DM models and their impact on structure formation see \citealt{Zavala2019}; for a review of the CDM challenges and possible solutions see \citealt{Bullock2017}). Regardless of the answer to these puzzles, the impact of the DM particle nature on the physics of galaxies remains a relevant factor that needs to be taken into account, not only because it causes a major and unavoidable uncertainty in structure formation, but also because the detailed properties of galaxies remain one of the most promising avenues to find clues about the DM identity. To incorporate new DM physics into structure formation theory, a novel framework has been proposed that
aims at mapping a broad range of DM particle physics models into a set of effective parameters that fully characterise structure formation at galactic scales \citep[ETHOS;][]{Cyr-Racine2016,Vogelsberger2016}.
Thus far, ETHOS covers two types of new DM physics:
(i) a primordial cutoff in the linear matter power spectrum suppressing the growth of small density perturbations due to either collisionless damping (free-streaming) like in WDM, or due to collisional damping caused by interactions between DM and relativistic particles in the early Universe and resulting in Dark Acoustic Oscillations (DAOs; for a review see \citealt{Bringmann2009}). DAOs are given explicitly in ETHOS by hidden sector DM-dark radiation (DR) interactions \citep{vdA2012,Buckley2014} but DM interactions with photons or neutrinos lead to a similar damping, \citep[e.g.][]{Boehm2002};
(ii) DM self-interactions (SIDM) reducing the central density of haloes in the non-linear regime. 

In this work we concentrate exclusively on (i) above, i.e., on the impact of a primordial DM cutoff with the objective of defining a parameter space that fully characterises structure formation within ETHOS at galactic scales (at high redshift $z>5$; see below). Ours is then a continuation of the work done in \citet{Cyr-Racine2016} where a small set of effective parameters was defined that where sufficient to characterise the \emph{linear} power spectrum in a variety of DM models with a cutoff. However, a large number of models that are different with respect to their linear power spectrum can in fact lead to identical structure formation. Therefore, we re-parametrize this effective ETHOS parameters, still being determined by the linear power spectrum, but with the goal of
providing a full account of the non-linear evolution of galactic-scale structures (down to $z=5$) using cosmological $N-$body simulations and a physical interpretation of the parameters; see Fig.~\ref{fig:cartoon}.  
We aim at dividing this new ETHOS parameter space into distinct structure formation regions, mapping smoothly between the different possibilities for the small-scale power spectrum (CDM, WDM or DAOs).

We note that previous works have proposed analytical formulae to describe the linear power spectra of different DM models, usually written as a transfer function relative to CDM \citep[e.g. for WDM][]{Bode2001,Viel2005,Leo2018}. More recently, \cite{Murgia2017,Murgia2018} proposed a formula for the transfer function that can seemingly accommodate WDM, fuzzy DM, and also certain ETHOS models. Crucially however, this formula does not describe DAOs since they were deemed not relevant for the properties of interest in \cite{Murgia2017,Murgia2018}, namely, for the 1D Lyman$-\alpha$ flux power spectrum, and for the number of observable Milky-Way subhaloes (i.e. those that can host a luminous satellite). As we demonstrate and quantify in this work, DAOs are quite relevant for a range of ETHOS models \citep[see also][]{Bose2019}. Moreover, our approach differs from previous ones since the parametrization we propose goes beyond providing a fit to the power spectrum, with the parameters having a clear physical interpretation.

In this work we study structure formation within ETHOS in the high-redshift Universe down to $z=5$. This choice was partly done to avoid entering the regime where DM self-interactions (another relevant ingredient in ETHOS) start to have a relevant impact in the centre of DM haloes. We are also only considering the impact of DM physics in structure formation without taking into account the role of baryonic physics, which clearly plays a role in DM clustering, albeit considerably smaller at high-redshift relative to the low-redshift Universe. In this way we can isolate the potential difference between CDM and other DM models, purely due to DM physics, without the influence of baryons; this is in fact needed to disentangle the impact of both effects. Our plan in future work is to extend the spirit of this work, by defining the space of structure formation parameters that are relevant for the physics of galaxies, to include both SIDM and baryonic physics.
We also choose $z=5$ as the lowest redshift we examine since it is roughly the maximum redshift where data from the Lyman-$\alpha$ flux-power spectra have been used to constrain the DM power spectrum at small scales \citep[e.g.][]{Viel2013,Murgia2018}. We use this both to exemplify how our parametrization can be used to potentially constrain DM models and to define the maximum scale where new DM physics can play a role in galactic-scale structure formation: DM models with a non-linear power spectrum significantly deviating from
CDM at $k\lesssim10$\hmpc are not compatible with the data (\citealt{Irsic2017}; although see \citealt{Garzilli2019}). 
On the other hand, we set the relevant minimum scale to be given by the atomic cooling limit (specifically, the primordial gas in haloes with a virial temperature $\lesssim 10^4$~K cannot cool via atomic transitions; see \citealt{White1978}). Galaxy formation is thus suppressed for DM haloes with masses below $\sim 10^8 {\rm M}_\odot\,h^{-1}$ (corresponding to non-linear scales of $\sim 500$\hmpc).
In summary, we study non-linear structure formation down to $z=5$ within the ETHOS framework using DM-only $N$-body simulations focusing on the non-linear scale range $10$\hmpc$\lesssim k\lesssim500$\hmpc~(halo virial masses in the range $10^8$M$_\odot\lesssim M_{\rm vir}\lesssim10^{10}$M$_\odot$).

This paper is organized as follows. In Section~\ref{sec:method}, we describe our simulation setup and the zoom-in method we use to cover the dynamic range of interest. The convergence properties of our simulations is discussed in Appendix~\ref{sec:convergence}. In Section~\ref{sec:parametrization}, the new ETHOS parametrization is constructed and connected to that in \cite{Cyr-Racine2016} (see also Appendix~\ref{sec:linpk}). In Section~\ref{sec:results}, we present our main results on how different structure formation models are classified within the new parametrization based on both the non-linear power spectrum and the halo mass function. Finally, our conclusions are given in Section~\ref{sec:conclusion}. 

\section{Numerical Methodology}
\label{sec:method}

The cosmological dark-matter-only $N-$body simulations used in this work were performed with the code \arepo~\citep{Springel2010}. Initial conditions for the simulations were generated using the code \music~\citep{Hahn2011} with the cosmological parameters set to $\Omega_{\rm m}=0.31069$, $\Omega_\Lambda=0.68931$, $H_0=67.5\,{\rm km/s/Mpc}$, $n_{\rm s} = 0.9653$ and $\sigma_8=0.815$, where $\Omega_{\rm m}$ and $\Omega_\Lambda$ are the contributions from matter and cosmological constant to the matter-energy density of the Universe today, respectively, $H_0$ is the Hubble constant today, $n_{\rm s}$ is the spectral index, and $\sigma_8$ is the mass variance of linear fluctuations in $8$\mpch spheres at $z=0$. This choice of cosmological parameters is consistent with a Planck cosmology \citep{Planck2018}. The linear power spectrum used as an input for \music~for the different DM models we explore is computed with a modified version of \classcode\footnote{\cite{Blas2011} (\href{http://class-code.net}{class-code.net})} \citep{Archidiacono2017,Archidiacono2019}\footnote{Note that in the first ETHOS paper \citep{Cyr-Racine2016}, this implementation is done with \camb~\citep{Lewis:2002ah}.}

For this work, we are interested in the matter power spectrum at the {\it non-linear} scales relevant for dwarf galaxies at high redshift from $k\sim10$ to $500$\hmpc. This range roughly corresponds to halo masses $\sim10^{10}$ to $10^{7}$~{\rm M}$_\odot$, which are close to the limits where significant deviations from CDM are possible, and galaxy formation becomes highly inefficient, respectively. Achieving a fair representation of the power spectrum and the halo mass function at such small scales was not feasible with a uniform simulation box due to the following stringent limitation.
At a fixed spatial resolution, reducing the size of the cosmological box, reduces the minimal scales probed, but at the cost of
missing the power transferred from larger to smaller scales in the non-linear evolution. Thus, the power spectrum at the scales and redshifts of interest would be biased towards lower values. This problem can be alleviated by having a sufficiently large simulation box, but in order to
resolve $500$\hmpc, the amount of particles and thus the computational cost increases dramatically. 
To achieve our goals we therefore rely instead on cosmological {\it zoom-in} simulations by using the method described in the following.

\subsection{Small-scale power spectrum with zoom simulations}\label{zoom_tec}

In a zoom-in cosmological simulation, the computational resources are focused on a smaller subregion within a large cosmological box. This subregion is simulated at the desired highest resolution, while the volume around contains low resolution elements that still preserve an accurate representation of large scale properties of the density field. The region of interest is usually a halo and its immediate environment, and the procedure to construct the initial conditions for zoom simulations consists of: i) run a low resolution {\it parent} cosmological simulation within a cosmological box large enough to provide a fair representation of the clustering properties at the scales of the box in the range of redshifts of interest (we found that a box size of 40\mpch per side satisfies this for $z\geq5$); ii) select a volume within the parent simulation encompassing the region of interest at the redshift of interest; iii) trace back the particles within this region to the starting redshift of the resimulation; this represents the target {\it Lagrangian} volume for the zoom simulation; (iv) finally, an initial conditions code like \music~is used to generate a multi-layered resolution coverage of the resimulation specifying the volume that covers this Lagrangian region with the highest resolution required. For more details of the general procedure see \citet{Onorbe2014}. 

\begin{figure}
    \centering
    \includegraphics[width=\columnwidth]{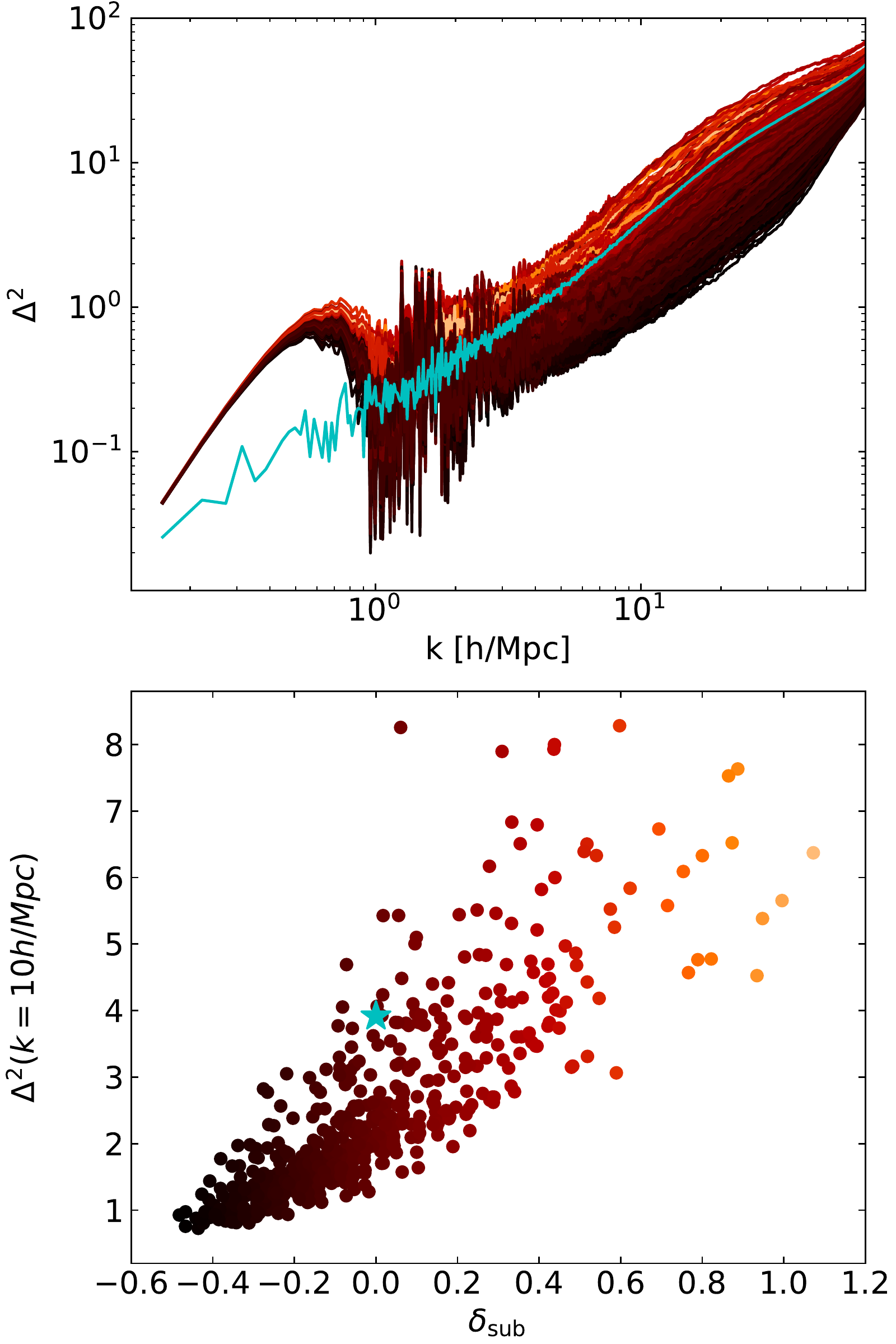}
    \caption{{\it Top panel}: Dimensionless power spectra $\Delta^2$ at $z=5$ of a large $40$\mpch cubic uniform simulation (cyan) and of sub-regions inside this simulation ($\sim 6.25$\mpch) coloured from under-dense (black) to over-dense (orange) relative to the larger box. The excess of power at small $k\lesssim1$\hmpc in the small box simulations is due to the finite size of the sub-regions, while at large $k$ the models start to converge artificially due to Poisson noise, which starts to dominate the signal (visible at around $k\gtrsim40$\hmpc). {\it Bottom panel}: Correlation of the dimensionless power spectrum at $10$\hmpc and the overdensity of the sub-region $\delta_{\rm sub}$ (the colour scale is as in the top). The cyan star corresponds to the value for the whole simulation box.}
    \label{fig:variations}
\end{figure}

In our case we followed the previous standard zoom-in procedure but not focusing on a particular halo in the parent simulation but rather on a smaller subregion with the main requirement for it to have a {\it similar} power spectrum compared to the larger parent box at the resolved scales. In the following we describe how we find the optimal sub-region according to this requirement. 

We find that at the scales of interest the power spectrum has roughly a similar shape, but with an amplitude that correlates strongly with the overdensity $\delta_{\rm sub}$ of the subregion, as can be seen in Fig.~\ref{fig:variations}. This type of {\it cosmic variance} is a well known effect that has been studied in the past, particularly in the linear regime where $\delta_{\rm sub}\ll 1$. In this regime, it is possible to correct for this bias by e.g. using the separate universe approach, where each subregion is treated as a separate universe with a different cosmology, in this case a universe with different background density \citep[see e.g.][]{Chiang2014,Li2014}.
As is clear from the bottom panel of Fig.~\ref{fig:variations}, the distribution of $\delta_{\rm sub}$ is broad, covering values that are clearly non-linear anymore. This is because
we are looking at
smaller scales where the impact of non-linear effects is stronger and the variance of $\Delta^2$ for regions with the same overdensity can be quite large, weakening the correlation between the amplitude of the power spectrum and the overdensity.
Instead of trying to generalize the separate universe approach into the non-linear case, 
we decided to carefully select our high resolution sub-region so that it has a power spectrum that is as similar as possible to the one of the larger lower resolution region, at the scales that both can resolve. In this way a correction becomes unnecessary. This is sufficient for our purposes since we are only interested in an average measure of the power spectrum down to small (galactic) scales, rather than in its variance. Nevertheless, Fig.~\ref{fig:variations} gives an impression of the (cosmic) variance to be expected in the power spectrum for survey volumes that are small $\lesssim10$~Mpc.   

\subsection{Performance of the zoom-in simulation technique}

As a benchmark test for the reconstruction ability and resource advantage of the method described above, we performed four CDM simulations in a (40\mpch)$^3$ volume down to $z=5$. The baseline is a uniform simulation with $1024^3$ particles, while the other three are zoom simulations where the low-resolution region corresponds to $512^3$ particles. The first of these has a (12.5\mpch)$^3$ zoom region with an effective resolution of $1024^3$ particles, the second one has a (6.25\mpch)$^3$ zoom region with an effective resolution of $2048^3$ particles, and the third one has a (6.25\mpch)$^3$ zoom region with an effective resolution of $4096^3$ particles.

\begin{figure}
    \centering
    \includegraphics[width=\columnwidth]{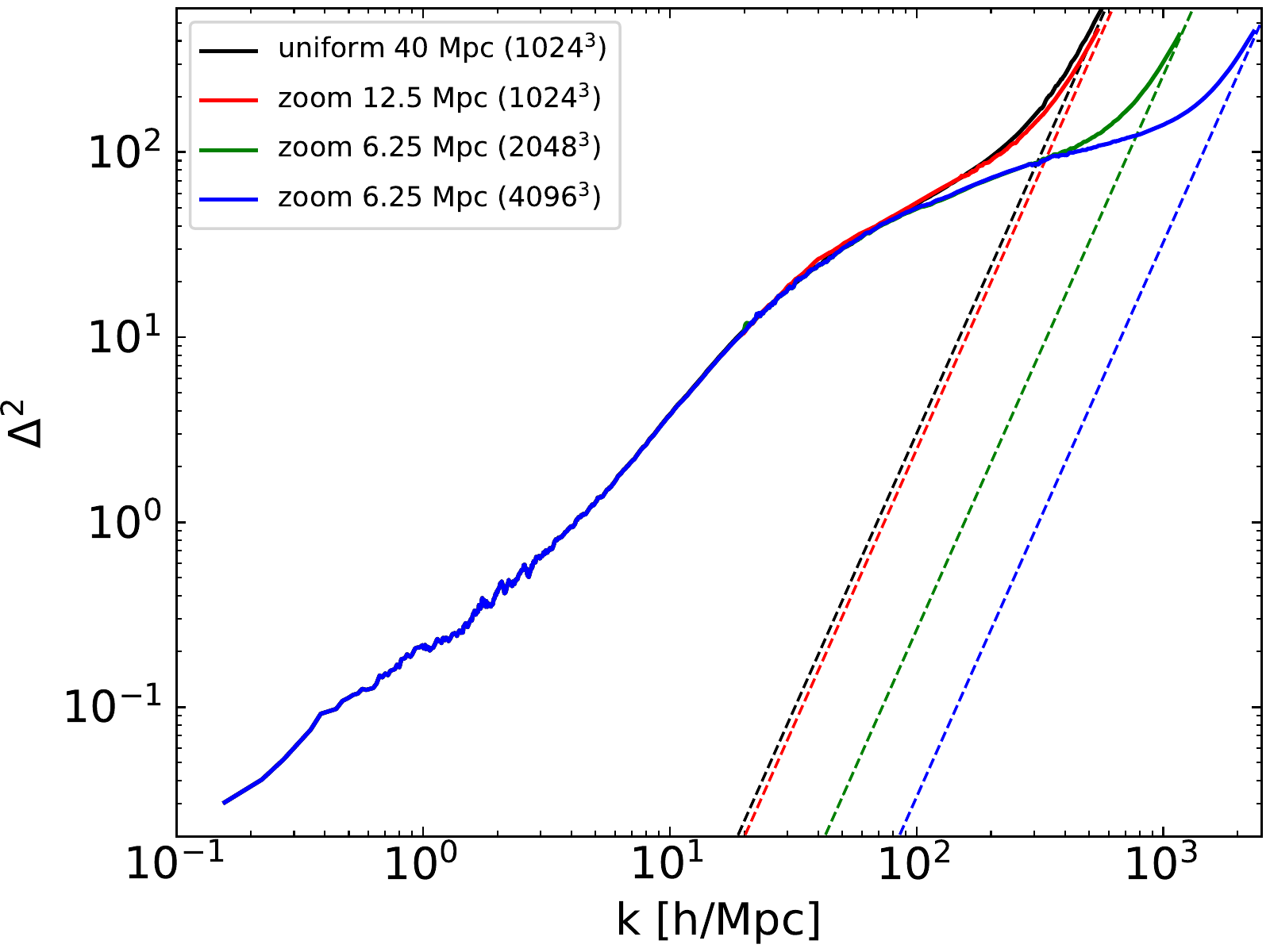}
    \caption{Dimensionless power spectra $\Delta^2$ at $z=5$ for a uniform simulation (black) and three zoom simulations with different zoom volumes and resolution levels. Note that the power spectrum from the zoom simulations can reach the same larger scales as the uniform simulation by including the low resolution particles. Thus, all lines completely overlap for $k\lesssim 30$\hmpc. The dashed lines show the expected Poisson (shot) noise ($\Delta^2_{\rm shot} = k^3 V / (2\pi^2N)$, where N is the number of particles and V the volume they occupy) for the corresponding simulation.}
    \label{fig:performance}
\end{figure}

In Fig.~\ref{fig:performance}, we can see that all three zoom simulations give a good reconstruction of the baseline power spectrum at large scales. As expected, the one with the same effective resolution as the baseline ($1024^3$ particles; red line) shows almost the same power spectrum as the uniform one at all scales, with nearly the same level of Poisson noise. The other two zoom simulations can resolve the power spectrum at smaller scales by factors of 4 (green) and 8 (blue) relative to the uniform simulation. This test shows that we can measure the power spectrum across a large dynamical range using the zoom simulation technique described in Section \ref{zoom_tec}. More importantly, it is possible to achieve this with a reduced computational cost as we show in Table \ref{tab:resources}.
For instance, the zoom simulation with same effective resolution as the uniform one (red and black lines in Fig.~\ref{fig:performance}) uses only a small fraction ($\lesssim1/7$) of the core hours and less than half the memory of the uniform simulation.
Even our highest resolution zoom simulation uses about the same core-hours and memory compared to the uniform run, while improving the scales that can be probed by a factor of 8, making it a very affordable approach to probing the power spectrum at small scales. 

\begin{table}
    \centering
    \begin{tabular}{c|c|c|c}
         & core-h & memory & resolved $k$ \\
        uniform ($1024^3$) & 14.9k & 594 GB & $\sim100$\hmpc \\ 
        \hline 
        12.5 Mpc ($1024^3$) & 2k & 247 GB & $\sim100$\hmpc \\ 
        \hline 
        6.25 Mpc ($2048^3$) & 2.9k & 259 GB & $\sim400$\hmpc \\ 
        \hline 
        6.25 Mpc ($4096^3$) & 15.1k & 529 GB & $\sim800$\hmpc \\ 
    \end{tabular}
    \caption{Computing resources needed to reach $z=5$ for a uniform simulation with $1024^3$ particles and three different zoom simulations.}
    \label{tab:resources}
\end{table}

Based on this test, we use the following setting for the results presented in this paper (unless stated otherwise): a cosmological box with 40\mpch on a side with a high resolution zoom region (effective resolution of 4096$^3$ particles with a particle mass of $8\times10^4\,{\rm M_\odot}h^{-1}$) covering a $\sim$(6.25\mpch)$^3$ Lagrangian volume, surrounded by a low resolution region (effective resolution of 512$^3$ particles), and intermediate resolution levels as a buffer zone between them. With this setting, we find that the power spectra of all DM models presented in this work is converged to better than $5\%$ at $500$\hmpc, while the halo mass function is converged to better than $5\%$ down to $10^8 {\rm M_\odot}/h$. In Appendix \ref{sec:convergence} we explicitly show the convergence tests we performed. 

\section{Parametrization of the linear power spectrum}
\label{sec:parametrization}

\begin{figure}
    \centering
    \includegraphics[width=\columnwidth]{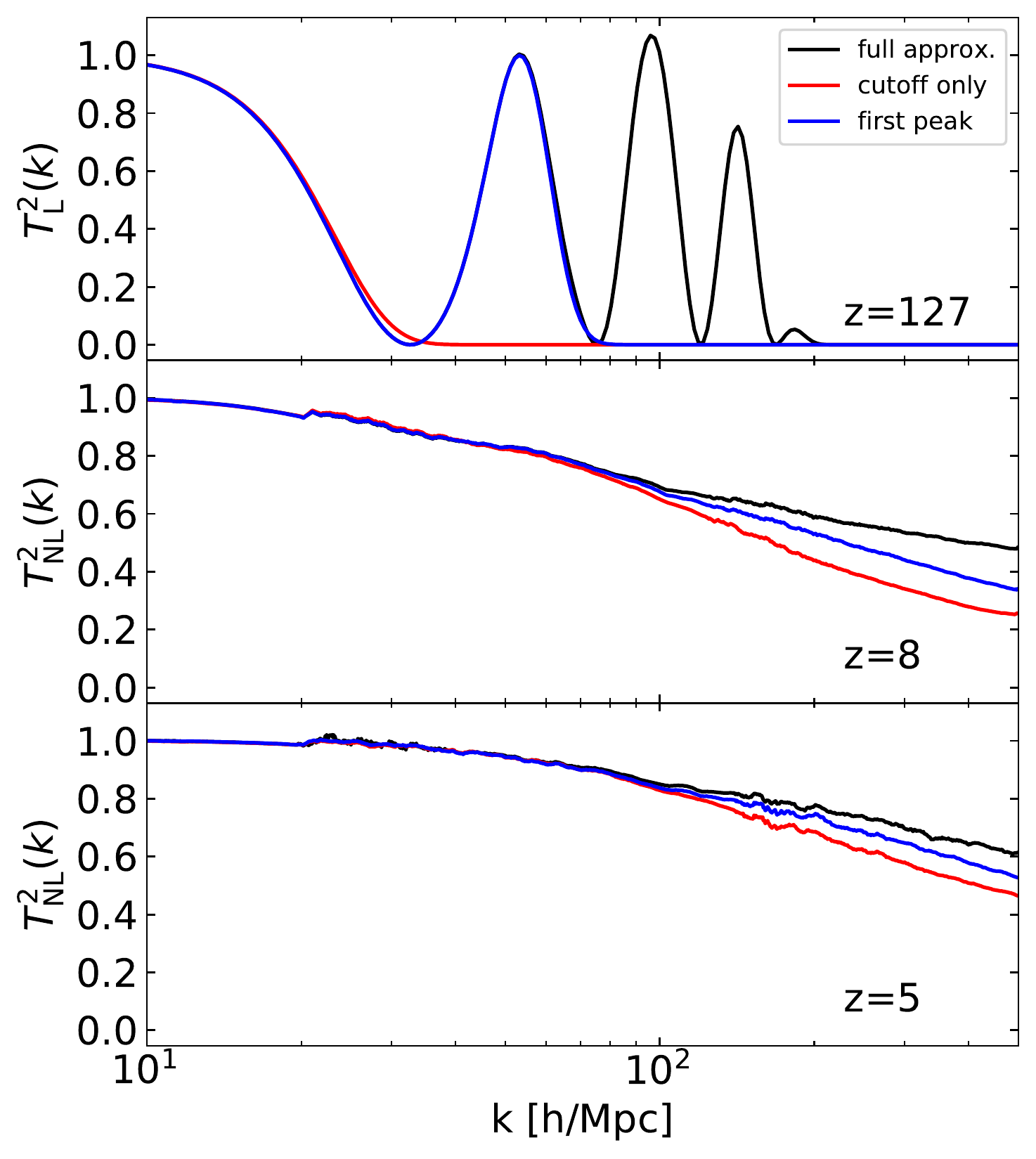}
    \caption{{\it Top panel}: Initial transfer function $T_{\rm L}^2(k)$ of a sDAO model computed with Eq.~\eqref{eq:parametrization} (black line) and two approximations: considering only the initial cutoff (i.e. with a WDM-like parametrization; see Eq.~\ref{eq:cutoff}), and adding as well the first oscillation (blue line; see Eq.~\ref{eq:gaussian}). {\it Middle and bottom panels}: Comparison of the non-linear dimensionless power spectra $\Delta^2$ relative to CDM at $z=8$ and $z=5$ for the models shown in the top panel.
    All these models used the following parameters: $h_{\rm peak}=1$, $k_{\rm peak} = 53.3$, $h_2 = 1.08$, $\tau = 0.67$, $\sigma = 0.2$, $\beta=4.05$, $\gamma=-20$,  $d = 2.5$, and $\alpha$ according to Eq.~\eqref{eq:alpha}. This corresponds to a model in the DAO region of Fig.~\ref{fig:result}.}
    \label{fig:param1}
\end{figure}

Our goal in this Section is to present a new parametrization of the linear power spectrum for DM models that have a primordial power spectrum cutoff with or without DAOs within the ETHOS framework. This parametrization is purely phenomenological but it is constructed with two objectives in mind: (i) although it parametrizes the linear power spectrum, its parameters should be sufficient to describe with good precision the non-linear power spectrum and (ii) the parameters should have a clear physical interpretation. To accomplish this, our starting point is the work of \citet{Murgia2017} who suggested the following parametrization for the cutoff of non-CDM (nCDM) models in terms of the linear transfer function $T_{\rm L}^2(k) \equiv P_{\rm nCDM}(k)/P_{\rm CDM}(k)$:
\begin{equation}
    T_{\rm L}(k) = [1+(\alpha k)^\beta ]^\gamma , \label{eq:cutoff}
\end{equation}
where $\alpha$ is a measure of the cutoff scale length, and $\beta$ and $\gamma$ the shape of the cutoff. This is a generalization of the fitting formula for WDM, where $\beta=2\nu$ and $\gamma=-5/\nu$ with $\nu=1.12$, and allows for much higher variety in the shape of the cutoff. However, it only describes a single cutoff in the power spectrum, while we want to include 
models with DAOs as well. 
We remark however that the transfer of power from large to smaller scales in the non-linear evolution tends to erase the DAOs \citep[e.g.][]{Buckley2014}. Since one of the goals of our parametrization is to reproduce with good precision the non-linear evolution of the power spectrum down to $z=5$, we thus start by looking at the accuracy to which Eq.~\eqref{eq:cutoff} can be expected to account for the non-linear regime.
This can be seen in Fig.~\ref{fig:param1}, where the red line corresponds to Eq.~\eqref{eq:cutoff}. Comparing the results from our simulations using this parametrization and the power spectrum with several DAOs (black lines) as initial conditions, we find that it is not sufficient to capture with precision the amplitude and features of the non-linear power spectrum at small scales for models which have strong DAO features (i.e. where the first oscillations are near the CDM amplitude).
As can be seen in Fig.~\ref{fig:param1} (red line), for this particular strong DAO model, this parametrization underestimates the power at $k\gtrsim100$\hmpc. For instance, by up to 48\% and 24\% at $k=500$\hmpc for $z=8$ and $5$ respectively. In Section \ref{sec:results_pk} we quantify in detail the impact of DAOs in the non-linear power spectrum for a broad range of scales and amplitudes of the DAOs.

In a first attempt to improve Eq.~\eqref{eq:cutoff} to account for DAO models, we add a term that includes the first DAO peak by modelling it with a Gaussian:
\begin{equation}
    T_{\rm L}(k) = [1+(\alpha k)^\beta]^\gamma - \sqrt{h_{\rm peak}}\exp \left( -\frac{1}{2}\left(\frac{k-k_{\rm peak}}{\sigma k_{\rm peak}}\right)^2 \right) , \label{eq:gaussian}
\end{equation}
where $h_{\rm peak}$ and $k_{\rm peak}$ give the amplitude (relative to CDM) and position of the first peak (these two will be the most important parameters throughout this work), and $\sigma$ controls how narrow the Gaussian is. Eq.~\eqref{eq:gaussian} improves the 
agreement with the full non-linear power spectrum 
as can be seen in Fig.~\ref{fig:param1}, but is still not good enough to reconstruct the full power at the smallest scales, where it underestimates the power by 30\% and 14\% for $z=8$ and 5 respectively. Therefore, the power provided by the secondary peaks in sDAO models remains relevant down to $z=5$.

To gain precision in our parametrization for models that have DAOs, we extend Eq.~\eqref{eq:gaussian} by adding terms that model the secondary peaks of the DAOs. These peaks can be described by two features, their envelope and oscillations. The oscillations of the higher order peaks are very regular and can be fitted with a cosine function, whose frequency is determined by $k_{\rm peak}$. The envelope can be parametrized with the amplitude of the second peak $h_2$ and two error functions, one giving the steep rise on the left side (similar to the Gaussian describing the first peak) and the other controlling the damping on the right (the oscillations are not fully symmetrical and thus a Gaussian is not enough to describe their shape). The full fitting function is then given by:

\begin{equation}
\begin{aligned}
    T_{\rm L}(k) &= [1+(\alpha k)^\beta]^\gamma - \sqrt{h_{\rm peak}}\exp \left( -\frac{1}{2}\left(\frac{k-k_{\rm peak}}{\sigma k_{\rm peak}}\right)^2 \right) \\
    &+ \frac{\sqrt{h_2}}{4} \operatorname{erfc} \left( \frac{k-1.805 k_{\rm peak}}{\tau k_{\rm peak}} - 2 \right) \\%\operatorname{erfc} \left(- \frac{k-1.805 k_{\rm peak}}{\sigma k_{\rm peak}} - 2 \right) \\
    &\times \operatorname{erfc} \left(- \frac{k-1.805 k_{\rm peak}}{\sigma k_{\rm peak}} - 2 \right) \cos \left( \frac{1.1083 \pi k}{k_{\rm peak}}  \right) ,
    \label{eq:parametrization}
\end{aligned}
\end{equation}
where $\operatorname{erfc}(x) = 1 - \operatorname{erf}(x)$ is the complementary error function. This full parametrization would have 8 parameters, but they are not all independent and can be simplified for ETHOS models, where they are fixed by $h_{\rm peak}$ and $k_{\rm peak}$ (see Section \ref{sec:connection} below). The parameter $\alpha$ can be determined by the scale at which the transfer function dropped to $1/2$ ($k_{1/2}$), which is connected to $k_{\rm peak}$: 

\begin{equation}
    \alpha = \frac{d}{k_{\rm peak}} \left[ \left(\frac{1}{\sqrt{2}}\right)^{1/\gamma} - 1 \right]^{1/\beta} ,\label{eq:alpha}
\end{equation}
where $d$ controls the ratio between $k_{\rm peak}$ and $k_{1/2}$, which is in the range $2.4-3$.

\begin{figure}
    \centering
    \includegraphics[width=\columnwidth]{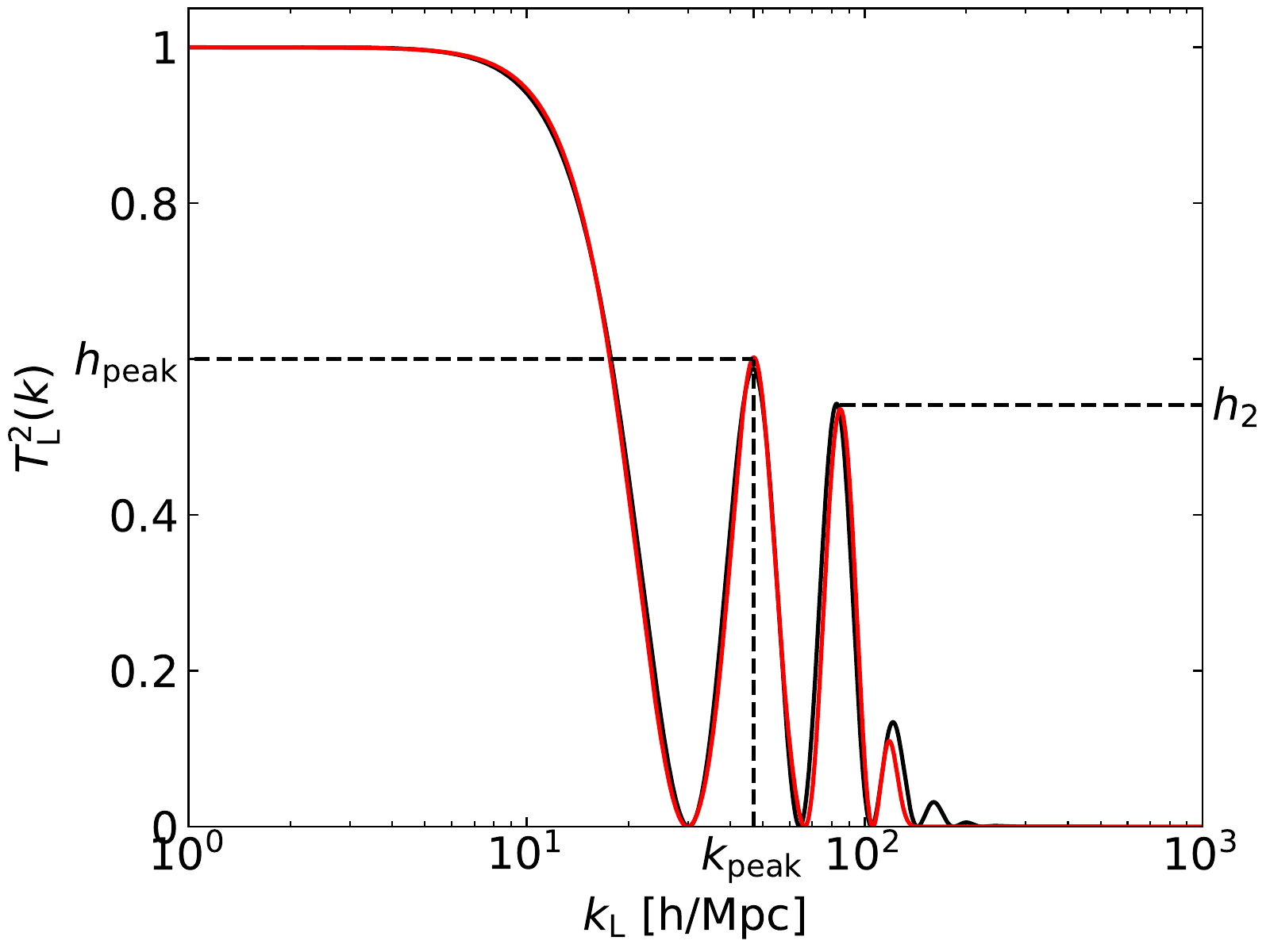}
    \caption{The transfer function $T_{\rm L}^2(k)$ of a DAO model computed with a Boltzmann solver (black) and fitted according to Eq.~\eqref{eq:parametrization} (red). The role of the most relevant parameters $h_{\rm peak}$, $k_{\rm peak}$ and $h_2$ is also shown.}
    \label{fig:parametrization}
\end{figure}

Fitting the power spectrum cutoff of ETHOS models with Eq.~\eqref{eq:cutoff} leads in all cases to large negative values for $\gamma$, whose precise value makes almost no difference in the reconstruction of the cutoff; thus, we have fixed $\gamma=-20$. From the remaining parameters, $h_{\rm peak}$ and $k_{\rm peak}$ are the most relevant parameters since the former determines the position of the first DAO peak as well as describing the position of the cutoff (see Eq.~\ref{eq:alpha}), and the latter the amplitude of the first DAO. These parameters are responsible for the leading order effects on the non-linear power spectrum and in fact the only free parameters within the models we study in this work. Regarding the remaining parameters: $\beta$ is responsible for the cutoff shape, $\tau$ controls the damping of the DAOs, and $\sigma$ gives the width of the first peak. Physically, $k_{\rm peak}$ is connected to the DM sound horizon and $\tau$ to the Silk damping scale (the physical interpretation of the key parameters is described in Section~\ref{sec:connection} and Appendix~\ref{sec:linpk}). The effects of the parameters and the quality of our final parametrization in the linear transfer function can be seen in Fig.~\ref{fig:parametrization}.

We emphasize that the parametrization given by Eq.~\eqref{eq:parametrization} can accurately describe the entire range of DM models in the ETHOS framework that display DAOs in their linear transfer function, including both weak and strong oscillations with only two free parameters ($h_{\rm peak},k_{\rm peak}$). Furthermore, it also naturally encompasses WDM ($h_{\rm peak} \rightarrow 0$) and CDM ($k_{\rm peak}\rightarrow \infty$), allowing us to explore a very broad range of possible DM physics.

%%%%%%%%%%%%%%%%%%%%%%%%%%%% parameter connection+used parameters. 
\subsection{Connection with the ETHOS framework and physical interpretation of the parameters: $h_{\rm peak}$, $k_{\rm peak}$ and $h_2$}\label{sec:connection}

Having accomplished the goal of providing a parametrization for DAO models that is simpler than a fully general parametrization of the linear power spectrum \citep[as provided in][]{Cyr-Racine2016} but still precise enough to describe their non-linear evolution, we proceed now to establish the connection between these phenomenological parameters and the physical parameters of the ETHOS framework \citep{Cyr-Racine2016} in regards to the effects of the DM-DR interactions in generating the power spectrum cutoff and the DAOs. We recall that with such a connection, it is then possible to have a complete mapping between the particle physics parameters of the models explored in \cite{Cyr-Racine2016} and the parameters relevant for non-linear structure formation. 

The physics of the DAOs in the linear power spectrum is captured within the modelling presented in \citet{Cyr-Racine2016}\footnote{We refer specifically to the case where DR-DR interactions are irrelevant.} by the parameters $n$ and coefficient $a_n$ that control the redshift scaling of the DM drag opacity $\dot\kappa_\chi \propto a_n (1+z)^{n+1}$, plus a set of coefficients $\alpha_l$ that parametrize the angular dependence of the DM-DR scattering cross section\footnote{More specifically, $\alpha_l$ is the ratio between the opacity of the $l^{\rm th}$-moment to that of the dipole moment of the DR multipole hierarchy given by the angular dependence of DM-DR scattering.}. For this work, we will refer only to models that have single values for $n$ and $a_n$, and a set of constant $\alpha_{l\geq2}$ values. A specific particle physics scenario contained within these constraints is that of a massive  fermionic  DM  particle interacting  with  a massless fermion via a massive vector mediator as in \cite{vdA2012}, which corresponds to the case $n=4$, $\alpha_{l\geq2}=3/2$ with different values of $a_4$ providing cutoff scales for the power spectrum. This specific model has been studied with simulations in the past \citep{Vogelsberger2016}, particularly the benchmark model referred to as ETHOS-4 in table 1 of \cite{Vogelsberger2016}.

Although the parameters $n$, $a_n$ and the set $\{\alpha_l\}$ are sufficient to characterize the linear power spectrum within the ETHOS framework, they obscure somewhat the physical mechanism behind the DAOs, and they also lack the simple phenomenological interpretation of the parameters described above $\{k_{\rm peak},h_{\rm peak},h_2\}$. Because of this, we first attempt to approximate the results of the full calculation of the linear power spectrum based on a Boltzmann code (modified version of \classcode; \citealt{Archidiacono2017,Archidiacono2019}) with a simple physical model based on the tight coupling limit approximation (between DM and DR) in analogy with the photon-baryon plasma (see e.g \citealt{Hu1996}). This attempt is described in Appendix~\ref{sec:linpk}. Although we find that this approximation is not accurate enough, particularly in describing the damping envelope of the DAOs, it does provide relevant insights into the relevance of the sound horizon scale and the DM decoupling epoch as the physical quantities behind the DAO features.
Therefore, we decided to try a phenomenological approach based on these quantities. To test this approach we explore a set of 84 ETHOS models as described above with the set of values: $\{n=(3-15,20),\log_{10}(a_n)=(0,1,2,3,5,7)\}$, and fixing $\alpha_{l\geq2}=3/2$. 

\begin{figure}
\centering
\includegraphics[width=\columnwidth]{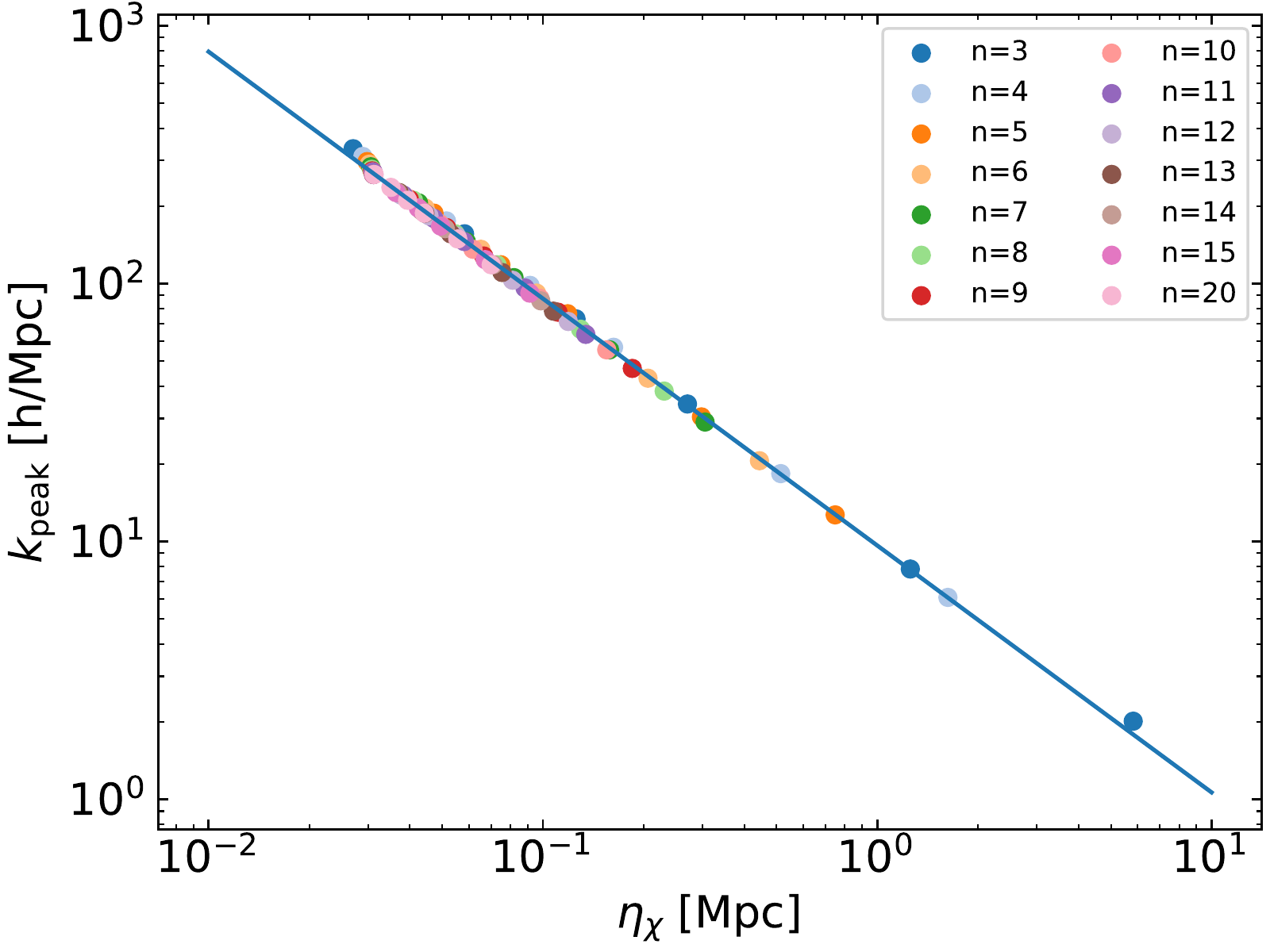}
\caption{The position of the first DAO peak $k_{\rm peak}$ correlates strongly with the time of DM decoupling $\eta_\chi$ defined by $\int_{\eta_\chi}^{\eta_0} -\dot\kappa_\chi d\eta=1$. Each symbol correspond to a different ETHOS model within a grid of $\{n,a_n\}$ values and fixing $\alpha_{l\geq2}=3/2$. Models with a fixed $n$ but different $a_n$ are represented with the same colour as given in the legend. The blue line is a power law fit to the correlation, $k_{\rm peak} = 9.37 \left(\eta_\chi/{\rm Mpc}\right)^{-0.97}$\hmpc.}
\label{fig:kpeak}
\end{figure}

The sound horizon scale $r_{\rm DAO} \approx c_{\rm s}\eta_\chi$, where $c_{\rm s}$ is the DM sound speed and $\eta_\chi$ is the conformal time of DM decoupling defined by 
\begin{equation}
    \int_{\eta_\chi}^{\eta_0} -\dot\kappa_\chi d\eta=1
\end{equation} 
with $\dot\kappa_\chi$ being the DM drag opacity due to the DM-DR interactions, should give the largest scale affected by acoustic oscillations and thus should be connected to $k_{\rm peak}$. We found this to be almost accurate, with only a slight deviation from a linear relation (see Fig.~\ref{fig:kpeak}): 
\begin{equation}
    k_{\rm peak} = 9.37 \left(\frac{\eta_\chi}{{\rm Mpc}}\right)^{-0.97} h\,{\rm Mpc^{-1}} ,
\end{equation}
i.e., $k_{\rm peak}$ is given by the sound horizon scale at the time of kinetic decoupling, with just a minor modification. Notice that since $a_n$ is connected to the decoupling time (see Eq.~\ref{eq:eta_DR} and \ref{eq:eta_chi}), then this relation implies a direct connection between $k_{\rm peak}$ and $a_n$. 

\begin{figure}
\centering
\includegraphics[width=\columnwidth]{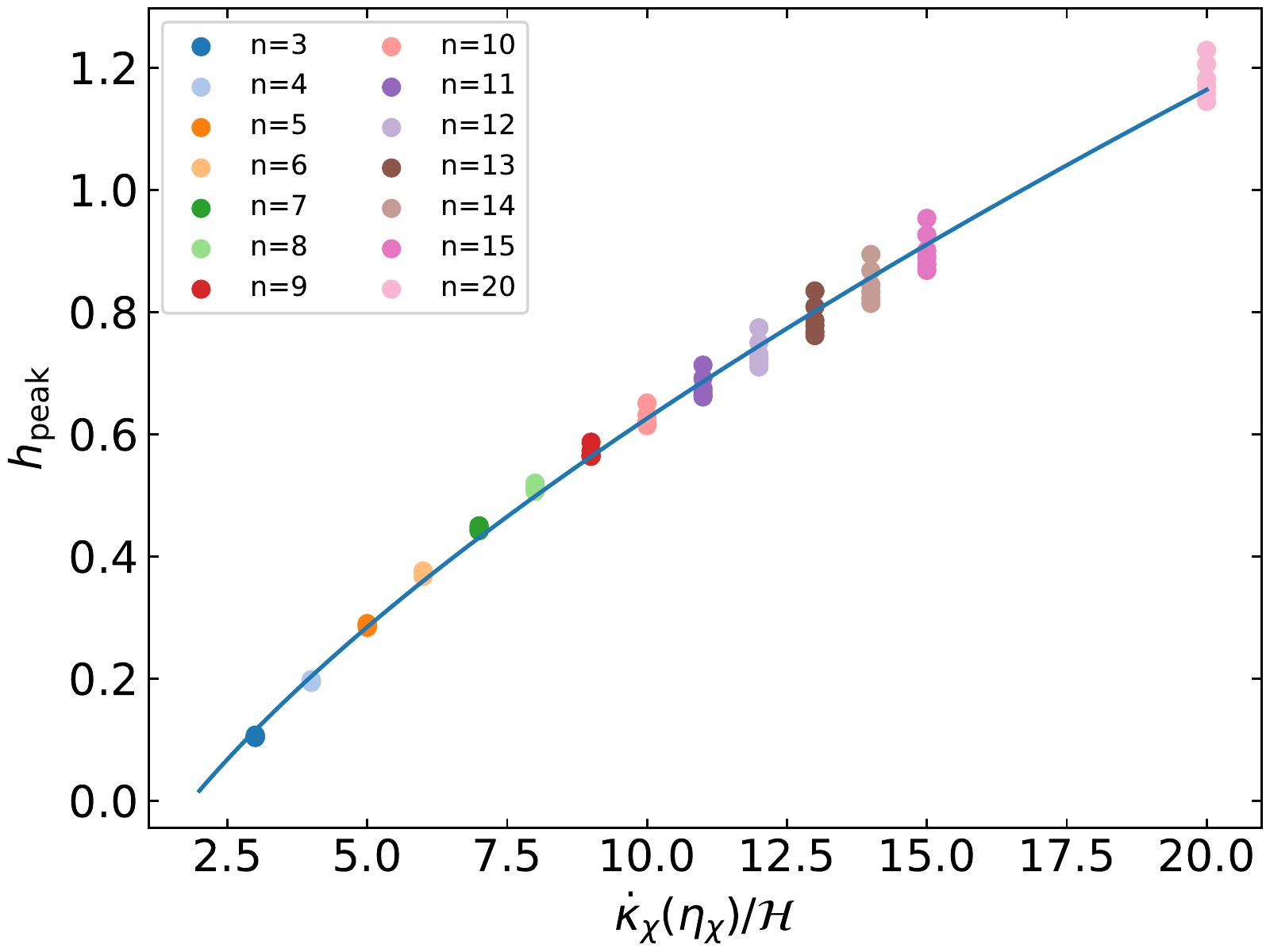}
\caption{The amplitude of the first DAO peak $h_{\rm peak}$ scales with the ratio of the DM drag opacity to the Hubble rate at the time of DM decoupling $\dot\kappa_\chi(\eta_\chi)/\mathcal{H}(\eta_\chi)=n$. The colours are the same as in Fig.~\ref{fig:kpeak}. The blue line is given by $h_{\rm peak} = 0.21 \left(\frac{\dot\kappa_\chi(\eta_\chi)}{\mathcal{H}(\eta_\chi)}\right)^{0.66} - 0.31$. }
\label{fig:h}
\end{figure}

On the other hand, we find that the damping of the first DAO, and therefore the parameter $h_{\rm peak}$, is mostly controlled by the DM mean free path (due to the DM-DR interactions) at DM decoupling, which is given by the inverse of $\dot\kappa_\chi(\eta_\chi)$ (see Fig.~\ref{fig:h}):
\begin{equation}
    h_{\rm peak} = 0.21 \left(\frac{\dot\kappa_\chi(\eta_\chi)}{\mathcal{H}(\eta_\chi)}\right)^{0.66} - 0.31 ,
\end{equation}
where the relevant quantity is actually the ratio $\dot\kappa_\chi(\eta_\chi)/\mathcal{H}(\eta_\chi)$, with $\mathcal{H}$ being the Hubble rate (relative to the conformal time). This ratio is actually equal to the ETHOS parameter $n$ (see Appendix~\ref{sec:linpk}). Thus, for $n\gg1$, the DM drag visibility function $\dot\kappa e^{-\kappa_\chi}$ is narrower, which implies a faster decoupling time scale; indeed, the DM-DR plasma is clearly in the tightly coupled regime at $\eta_\chi$, and thus the damping by DR diffusion is only significant for the smallest scales. On the contrary when $n\gtrsim1$, the DM drag visibility function is broader so that the timescales for decoupling (which occurs mostly in the weakly coupled regime) are larger. Thus, the DM mean free path at decoupling is relatively large and DM can diffuse substantially, lowering the value of $h_{\rm peak}$. While for $n\lesssim 9$, $n$ is the only factor in determining $h_{\rm peak}$, that is not true anymore for $n\geq 10$. In the latter case, $h_{\rm peak}$ spreads around the fit in Fig.~\ref{fig:h} depending on the value of $a_n$, the larger $a_n$ the larger $h_{\rm peak}$.  

\begin{figure}
\centering
\includegraphics[width=\columnwidth]{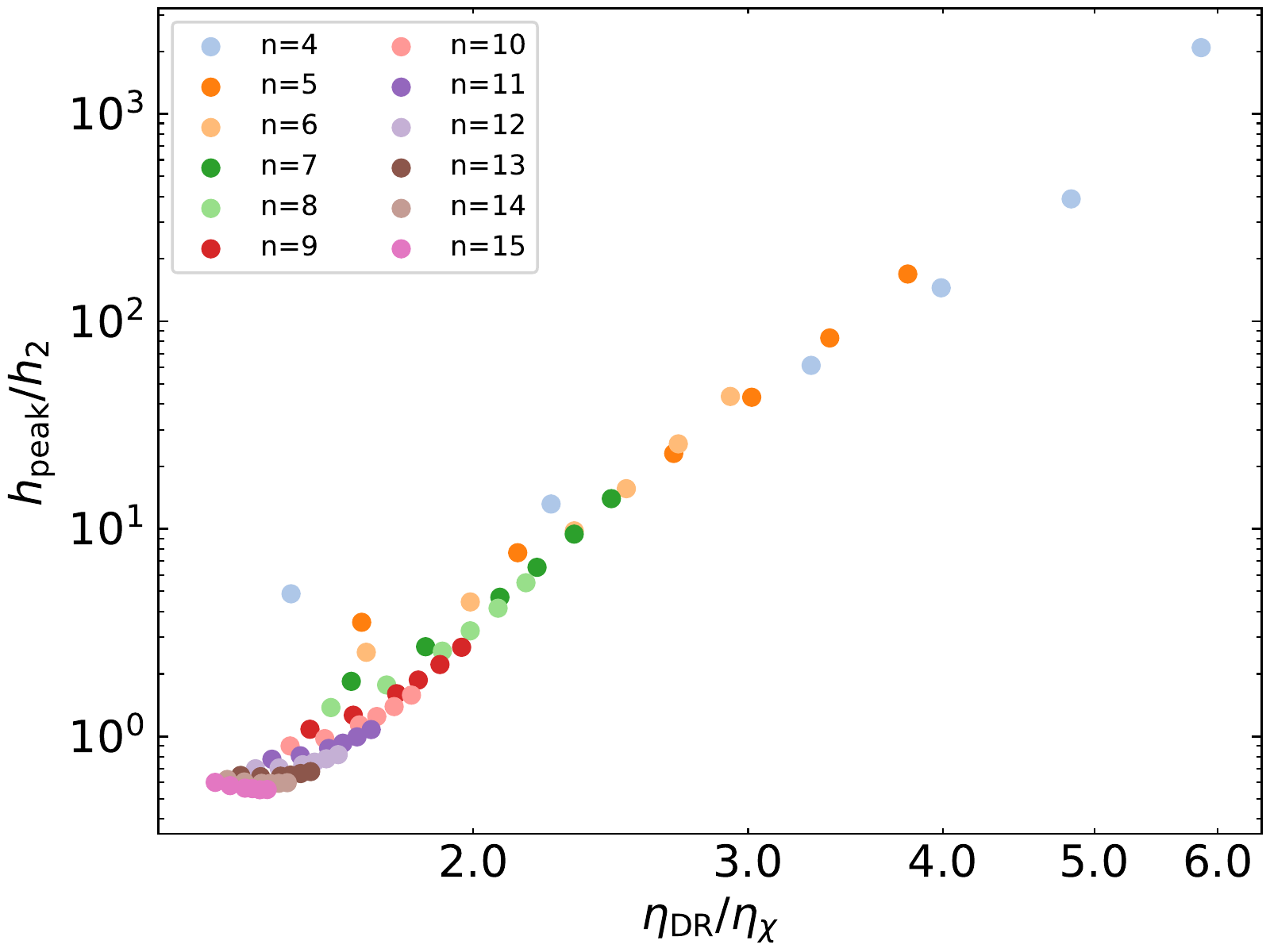}
\caption{The ratio of the first two DAO peaks scales with the ratio of the DR and DM decoupling times. The relation is nearly a parabola, with the parameters depending on the specific value of $n$. The colours are the same as in Fig.~\ref{fig:kpeak}.}
\label{fig:tau}
\end{figure}

Finally, the ratio of the first two DAO peaks $h_{\rm peak}/h_2$, is connected to the ratio of the DR to DM decoupling times $\eta_{\rm DR}/\eta_\chi$ (see Fig.~\ref{fig:tau}), where $\eta_{\rm DR}$ is the conformal time of DR decoupling defined by 
\begin{equation}
    \int_{\eta_{\rm DR}}^{\eta_0} -\dot\kappa_{\rm DR} d\eta=1
\end{equation} 
with $\dot\kappa_{\rm DR}$ being the DR opacity to DM scattering. This relation can be approximated by a parabola, but with different parameters for different values of $n$.

There are two additional features in Fig.~\ref{fig:h} that we highlight: (i) for models with very large $n\gtrsim15$ it is possible to have $h_{\rm peak}>1$ \citep[see also][]{Kamada2018,Ando2019}; (ii) for models with $n\gtrsim10$, the value of $h_{\rm peak}$ depends not only on $n=\dot\kappa_\chi(\eta_\chi)/\mathcal{H}(\eta_\chi)$, but also on the specific value of $a_n$ (the larger $n$, the stronger the dependence). For the latter models it is also true that the second DAO peak is significantly larger than the first (see Fig.~\ref{fig:tau}). We decide to exclude these models, i.e. those with $n\gtrsim10$, from our analysis for two reasons: (i) they would have a power spectrum that exceeds that of CDM at certain scales, and (ii) the parameter $h_2$ would no longer be a secondary parameter in determining the non-linear power spectrum.
We notice however, that there is potentially interesting phenomenology in these models, which we leave for a future work.
With this exclusion and using the strong correlations seen in Figs.~\ref{fig:kpeak}-\ref{fig:tau}, we have accomplished our goals at least for the regime of weak to moderately strong ($h_{\rm peak}\sim0.6$) DAO models, i.e., we have found a way to connect the parameters $h_{\rm peak}$ and $k_{\rm peak}$ in our parametrization to $n$ an $a_n$, respectively, in the original ETHOS framework, as well as to connect $h_{\rm peak}$, $k_{\rm peak}$ and $h_2$  to physical quantities that are responsible for the DM-DR decoupling.

\begin{figure}
\centering
\includegraphics[width=\columnwidth]{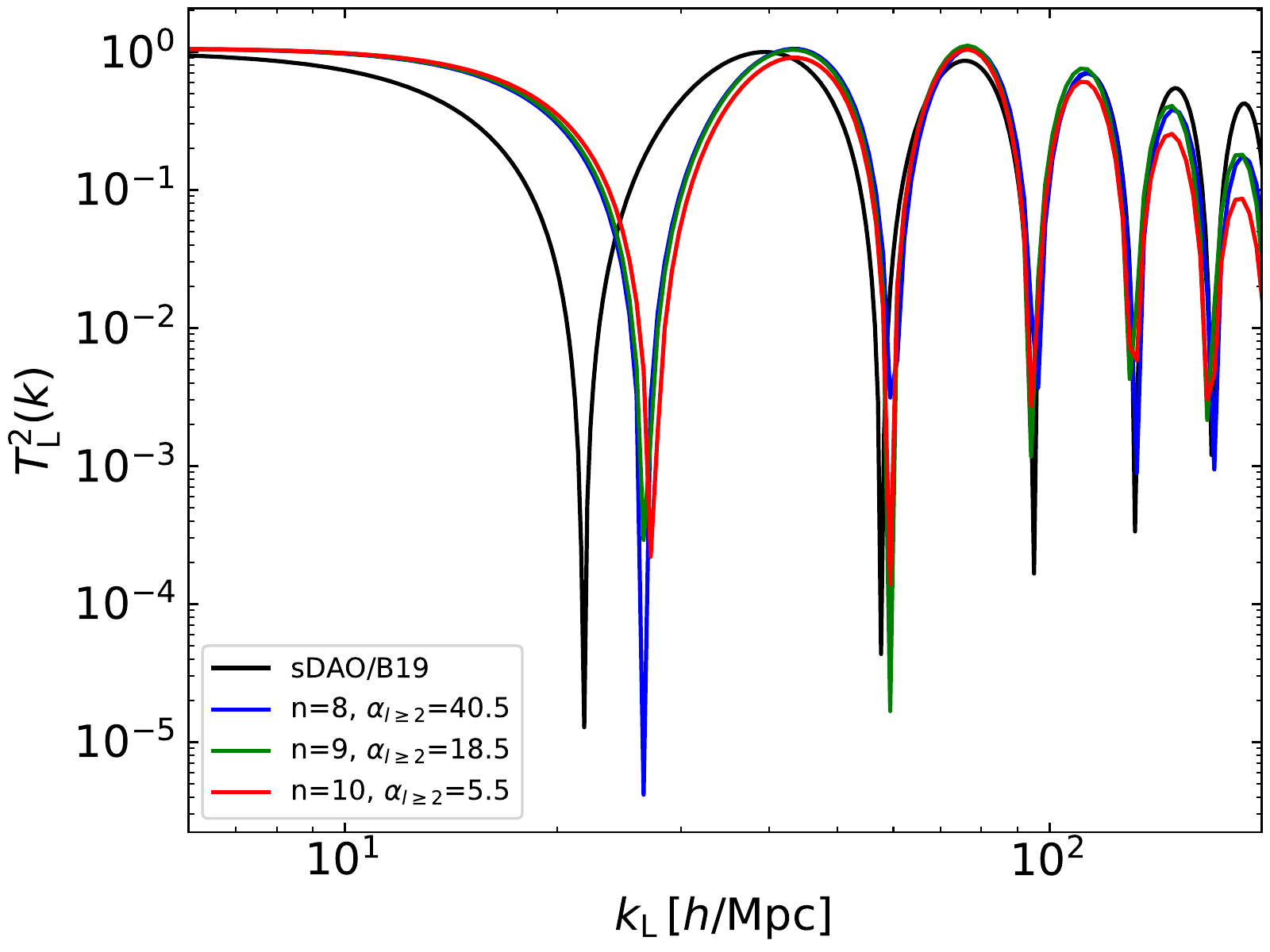}
\caption[]{Power spectra of strong DAO models where the first few DAO peaks have roughly the same height, reaching the CDM amplitude. The black line is the sDAO model from \cite{Bose2019}, while the lines with different colours are for ETHOS models. It is possible to accomplish this behaviour within the ETHOS framework for a fixed value of $n$ by systematically changing the value of $\alpha_{l\geq2}$ for $n=8,9,10$.} %add [] after caption to remove error message
\label{fig:aggressive}
\end{figure}

Within our parametrization, it is possible to include models with stronger DAO features ($h_{\rm peak}\sim1$), but that do not exceed greatly the CDM power spectrum, such as the benchmark sDAO model analysed in \cite{Bose2019} to show the distinct features this type of models leave in the Ly-$\alpha$ forest 1D flux spectrum. To do so, we need to change the value of $\alpha_{l\geq2}$ from $3/2$ to a value of $\mathcal{O}(10)$ for $n\sim9$. In this way, we can create power spectra that have strong DAO features but without having $h_{\rm peak}/h_2<1$ (see Fig.~\ref{fig:aggressive}).
This modification breaks the relations for the peak heights in Fig.~\ref{fig:h} and \ref{fig:tau}, while the one for $k_{\rm peak}$ stays unchanged. We used these models with increased $\alpha_{l\geq2}$ for our strong DAO cases.

\subsection{Final parameter space}\label{sec:parameterspace}

Given all previous considerations, we work with a 2D parameter space  with a set $\{k_{\rm peak},h_{\rm peak}\}$. We explore simulations within a range of  
$k_{\rm peak}$ between 35-300\hmpc and $h_{\rm peak}$ from 0 to 1. Notice that $h_{\rm peak}=0$ corresponds to thermal WDM models. This parameter space is covered with a grid of 52 simulations, spaced by 0.2 intervals in $h_{\rm peak}$ and equidistant in log$(k_{\rm peak})$ on two separate intervals: [35,100]\hmpc and [100,300]\hmpc. The parameters of our full parametrization (Eq.~\eqref{eq:parametrization}; Table~\ref{tbl:parameters}) have been calibrated to the linear power spectra of the corresponding ETHOS model obtained with the Boltzmann solver \classcode. We find that for all models, the parameters $\tau, \sigma, \beta$ and $d$ only depend on $h_{\rm peak}$.
Following the results in Fig.~\ref{fig:h}, we can relate $n=4$ for $h_{\rm peak}=0.2$, $n=6$ for $h_{\rm peak}=0.4$, and $n=9$ for $h_{\rm peak}=0.6$; all of these with constant values $\alpha_{l\geq2}=3/2$. The height of the second DAO peak, $h_2$, depends on $k_{\rm peak}$ (through the correlation seen in Fig.~\ref{fig:tau}) for all these models, but we find that can be modeled with a simple exponential function $h_2=A\exp{(Bk_{\rm peak})}+C$ (see Table \ref{tbl:parameters} for the values of A, B, and C). For $h_{\rm peak}=0.8,1.0$, $n$ was fixed to 9, but $\alpha_{l\geq2}$ had to be increased to a value in between $\sim10-30 (30-100)$ for $h_{\rm peak}=0.8 (1.0)$, depending on the value of $k_{\rm peak}$, in order to reach the desired value of $h_{\rm peak}$ without having a very dominant second DAO peak, which in this case is independent of $k_{\rm peak}$. The final parameters used in our simulations for a given $h_{\rm peak}$ are given in Table~\ref{tbl:parameters}.
\begin{table}
    \begin{tabular}{|c|c|c|c|c|c|}
        \hline
        $h_{\rm peak}$ & $h_2$ & $\tau$ & $\sigma$ & $\beta$ & $d$ \\
        \hline
        0.0 & 0.0 & 0.0 & 0.0 & 2.24 & 3.0 \\
        \hline
        0.2 & (0.067,-0.086,0.0011) & 0.34 & 0.23 & 3.1 & 2.93 \\
        \hline
        0.4 & (0.221,-0.025,0.0129) & 0.34 & 0.22 & 3.61 & 2.61 \\
        \hline
        0.6 & (0.572,-0.008,0.1490) & 0.38 & 0.2 & 3.91 & 2.44 \\
        \hline
        0.8 & 0.88 & 0.55 & 0.2 & 4.0 & 2.46 \\
        \hline
        1.0 & 1.08 & 0.67 & 0.2 & 4.05 & 2.5 \\
        \hline
    \end{tabular}
    \caption{Parameters used to construct the linear power spectra used in our simulations (Eq.~\ref{eq:parametrization}) as a function of the amplitude $h_{\rm peak}$ of the first DAO peak relative to CDM. For $h_{\rm peak}$ in the range $[0.2,0.6]$, the amplitude of the second DAO peak, $h_2$ is given by $h_2=A\exp{(Bk_{\rm peak})}+C$, where the $(A,B,C)$ values are given in the column.}
    \label{tbl:parameters}
\end{table}
%%%%%%%%%%%%%%%%%%%%%%%%%%%%

\begin{figure*}
    \centering
    \includegraphics[width=\textwidth]{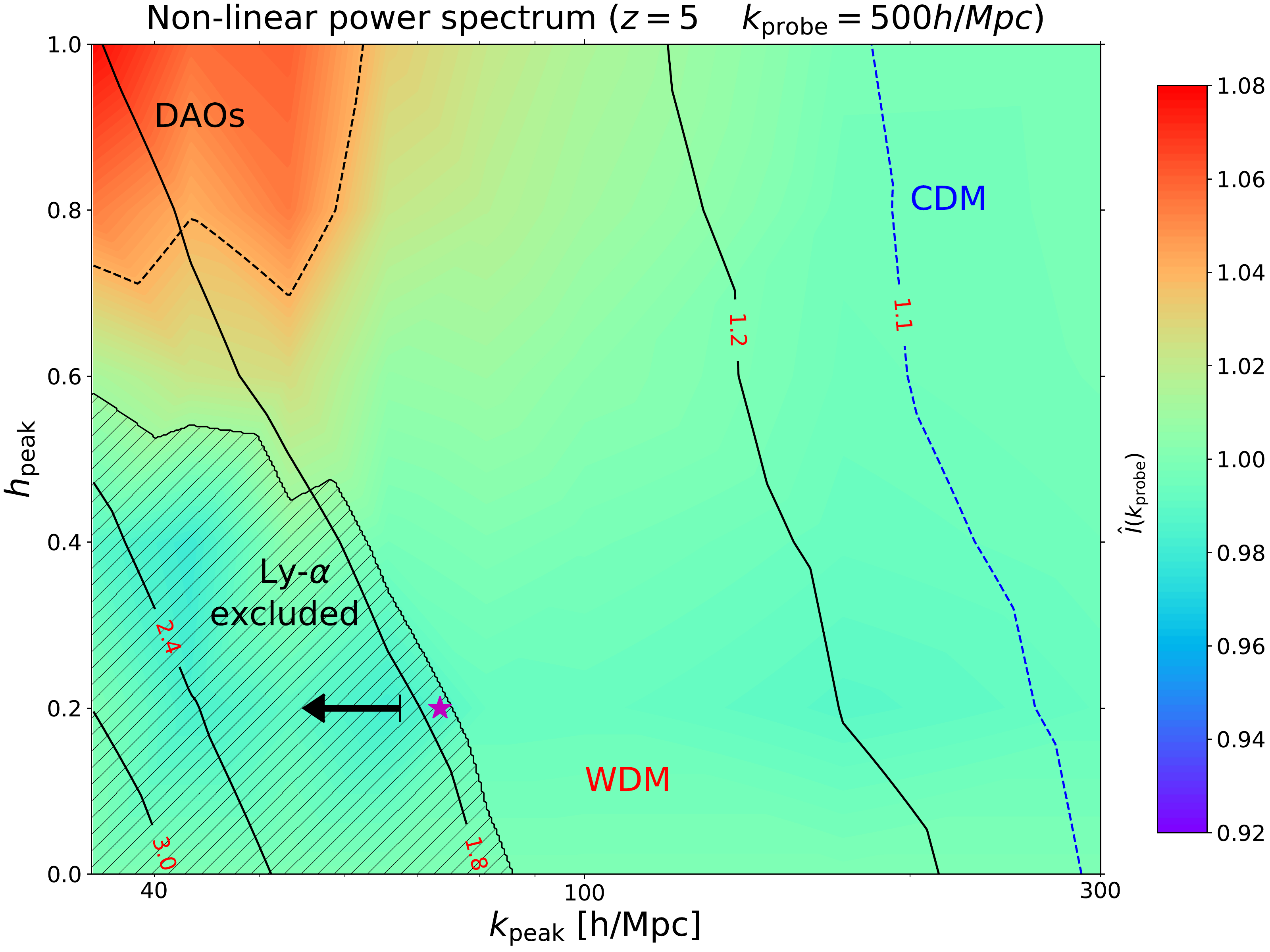}
    \caption[]{Division of structure formation models in the effective parameter space $h_{\rm peak}$ and $k_{\rm peak}$ (see Eq.~\ref{eq:parametrization}) according to their power spectra at $z=5$ for $k_{\rm probe}=500$\hmpc. The black contour lines correspond to the ratio $R(k_{\rm probe})$ of the CDM power spectrum relative to a given model (see Eq.~\ref{eq:ratio}). The colour scale shows the re-normalized values of the integrated quantity $\hat{I}(k_{\rm probe})$ (see Eqs.~\ref{eq:differentiator}$-$\ref{eq:iprobe2}), where a value of 1 corresponds to areas that are degenerate with WDM: models with $\hat{I}=1$ have the same power spectrum at $z=5$ as the WDM model at the same contour value of $R$. The black dashed line ($\hat{I}(k_{\rm probe})>1.04$) on the upper left encompass the area where DAO features survive until $z=5$. The region to the right of the blue dashed line, $R(k_{\rm probe})=1.1$, can be considered as nearly indistinguishable from CDM up to $k_{\rm probe}$. The hashed region on the lower left encompasses the area of models that are degenerate with a thermal WDM of mass <$3.6~{\rm keV}$, which has been ruled out by Lyman-$\alpha$ forest data \citep{Murgia2018}, and the arrow indicates the upper bound from the Lyman-$\alpha$ analysis of $h_{\rm peak}=0.2$ ($n=4$) models in \citet{Archidiacono2019}. The ETHOS-4 model used as a benchmark in \cite{Vogelsberger2016} is indicated by the purple star.}
    \label{fig:result}
\end{figure*}

\section{Results}\label{sec:results}

To characterise the differences between DM models and find out which features survive the non-linear evolution, we look at the matter power spectrum and the halo mass function at high redshift $z>5$.

\subsection{Matter power spectrum}\label{sec:results_pk}

We evaluate the power spectrum of the simulations at different scales $k_{\rm probe}$ between $10$\hmpc and $500$\hmpc. This range roughly covers the relevant range where new DM physics can play a role in the physics of galaxies: the larger scales are bounded by current constraints over deviations from CDM (for instance from Ly-$\alpha$ forest measurements e.g. \citealt{Irsic2017}), while the smaller scales are bounded by the minimum scales at which galaxies can form, where we use the atomic cooling limit as a reference. At $z=5$, the virial mass of a halo corresponding to a non-linear scale of $k=500$\hmpc is $\sim 5\times10^7\,{\rm M}_\odot\,h^{-1}$, which is just below the atomic cooling limit at this redshift, $\sim 10^8\,{\rm M}_\odot\,h^{-1}$. 

To quantify the difference of a given model with respect to CDM, we define two different diagnostics: (i) we compute the ratio $R(k_{\rm probe})$ of the power spectra with respect to CDM at $k_{\rm probe}$: 
\begin{equation}\label{eq:ratio}
    R(k_{\rm probe})=\frac{\Delta^2_{\rm CDM}(k_{\rm probe})}{\Delta^2(k_{\rm probe})},
\end{equation}
and consider $R(k_{\rm probe})\leq1.1$ as essentially indistinguishable from CDM (since we set our convergence goal to $5\%$; see Appendix~\ref{sec:convergence}); (ii) to distinguish different non-cold DM models, $R(k_{\rm probe})$ is not sufficient enough, as two models with the same ratio at a given $k$ can have different behaviour on larger scales. To capture this with a single number, we define the following dimensionless integrated quantity\footnote{Note that we use a quadratic dependence on the ratio $1/T^2_{\rm NL}=\Delta^2_{\rm CDM}/\Delta^2$ in the integrand in Eq.~\eqref{eq:differentiator} instead of a linear one in order to enhance the difference between models.}
\begin{equation}
    I(k_{\rm probe}) = \frac{\displaystyle{\int_{k_{\rm min}}^{k_{\rm probe}}} \left(\frac{\Delta^2_{\rm CDM}(k)}{\Delta^2(k)}\right)^2 d\ln{k}}{\ln{(k_{\rm probe}/k_{\rm min}})} ,
    \label{eq:differentiator}
\end{equation}
where we choose $k_{\rm min}=10$\hmpc since, as we mentioned above, the models we are interested in have the same power at this scale. 
By construction, a larger value of $R(k_{\rm probe})$ also results in a larger value of $I(k_{\rm probe})$, thus we need to normalize it to a reference case in order to define a comparative quantity across the different structure formation models. We choose the WDM case as the reference and normalize
Eq.~\eqref{eq:differentiator} for a given model by the value of $I(k_{\rm probe})$ of a WDM model with the same value of $R(k_{\rm probe})$:
\begin{equation}\label{eq:iprobe2}
    \hat{I}(k_{\rm probe})=\left(\frac{I(k_{\rm probe})}{I_{\rm WDM}(k_{\rm probe})}\right)_{R(k_{\rm probe})}
\end{equation}
Defined in this way, for a fixed $R(k_{\rm probe})$, all models with $\hat{I}(k_{\rm probe})=1$ have a non-linear power spectrum at the given redshift which is essentially indistinguishable from a WDM of the same $R(k_{\rm probe})$, regardless of how different the linear power spectrum of these models is relative to WDM.

Figure~\ref{fig:result} shows the results for our simulations for $z=5$ and $k_{\rm probe}=500$\hmpc in the leading order space of parameters $k_{\rm peak}$ and $h_{\rm peak}$ (Eq.~\ref{eq:parametrization}). We recall that we ran 50 simulations for models within this parameter space, which are then used to bilinearly interpolate the values of $R(k_{\rm probe})$ and $\hat{I}(k_{\rm probe})$ between the simulated models (the grid described in Section~\ref{sec:parameterspace}) to fill in Fig.~\ref{fig:result}. 
The line contours show $R(k_{\rm probe})$, which increases from right to left with the models to the right of the blue dashed line ($R=1.1$) being virtually indistinguishable from CDM, while those to the left become ever more divergent from CDM. 
The colour scale shows the value of $\hat{I}(k_{\rm probe})$ and therefore quantifies how different a model is compared with a WDM model that has the same value of $R$. Note that the WDM models in this plot lie at the bottom $k_{\rm peak}$ axis ($h_{\rm peak}=0$). Using the connection between $\alpha$ and $m_{\rm WDM}$ from \citet{Viel2005}
\begin{equation}
    \alpha = 0.049 \left(\frac{m_{\rm WDM}}{1\,{\rm keV}}\right)^{-1.11} \left(\frac{\Omega_\chi}{0.25}\right)^{0.11} \left(\frac{h}{0.7}\right)^{1.22} h^{-1}\,{\rm Mpc} ,
\end{equation}
we can compute $m_{\rm WDM}$ from $k_{\rm peak}$:
\begin{equation}
    \frac{m_{\rm WDM}}{1\,{\rm keV}} = \left[0.050 \left(\frac{k_{\rm peak}}{h\,{\rm Mpc}^{-1}}\right) \left(\frac{\Omega_\chi}{0.25}\right)^{0.11} \left(\frac{h}{0.7}\right)^{1.22} \right]^\frac{1}{1.11} 
\end{equation}
For a value of $\hat{I}$ close to one (green colour), the model's power spectrum (up to $k_{\rm probe}=500$\hmpc and at $z=5$) will be indistinguishable from WDM, while larger values mean that the power spectrum shape for $k<k_{\rm probe}$ is truly distinct from WDM regardless of the value of $R(k_{\rm probe})$. The region in the top left where $\hat{I}(k_{\rm probe})$ has the largest values corresponds to models with strong DAO features ($h_{\rm peak}\gtrsim0.7$ and $k_{\rm peak}\lesssim65$\hmpc; labeled as DAOs) where the impact of the DAO features has not been erased by the non-linear evolution down to $z=5$, and thus still leaves a signature in the power spectrum up to $k_{\rm probe}$. As is apparent most of the parameter space outside of the latter DAO region has values of $\hat{I}$ close to 1, and thus any models here are essentially degenerate (up to $k_{\rm probe}=500$\hmpc and at $z=5$) with a WDM model with the same value of $R$.  
This degeneracy is either caused by the non-linear evolution erasing the DAO features, especially for the weak DAO models in the lower part of the plot ($h_{\rm peak}\lesssim0.6$), or because the DAO features appear at smaller scales $k>k_{\rm probe}$ than we are interested in (for models with $k_{\rm peak}\gtrsim100$\hmpc). 
We remark that comparisons between WDM, wDAO and sDAO models have been done in the past using $N$-body simulations %, particularly to study if they look the same at large scales 
\citep[see e.g.][]{Buckley2014,Vogelsberger2016,Schewtschenko2015}. In particular, \citet{Murgia2018} have shown that the presence of weak oscillations does not affect the scales probed by Lyman-$\alpha$ forest observations. However, we are showing with Fig.~\ref{fig:result} that the degeneracies between DAO and WDM models extend to much smaller scales (including strong oscillations), and crucially, we introduce a quantity $\hat{I}(k_{\rm peak})$ that is a measure of the degree of degeneracy.

\begin{figure}
    \centering
    \includegraphics[width=\columnwidth]{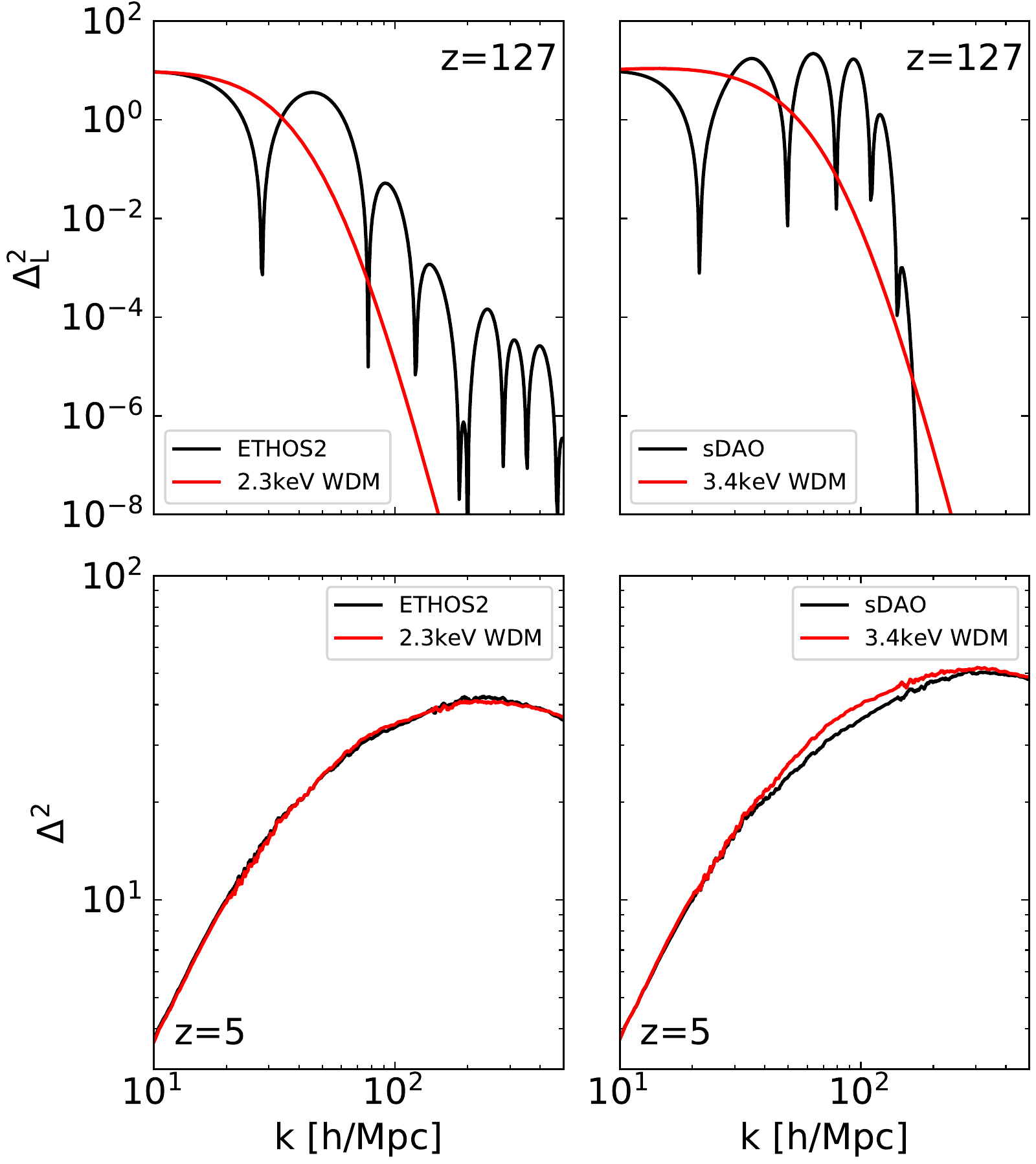}
    \caption{Comparison of the power spectra from a couple of DAO models and the comparable WDM models with $m=2.3\,{\rm keV}$ (left) and $m=3.4\,{\rm keV}$ (right) at the initial conditions for our simulations ($z=127$ top panels) and at the end of the simulations ($z=5$; bottom panels). The left panels are for a wDAO model (ETHOS-2 in \citealt{Vogelsberger2016}), while the right panels are for an sDAO model. On the left, the different models have clearly different power spectra at the initial conditions, but become completely degenerate at $z=5$. On the right panels on the other hand, the models remain different even at $z=5$, despite having the same power at $k=500$\hmpc. This shows that only strong DAO models have truly distinct power spectra features relative to the WDM model.}
    \label{fig:degen}
\end{figure}

\begin{figure*}
    \centering
    \includegraphics[width=\textwidth]{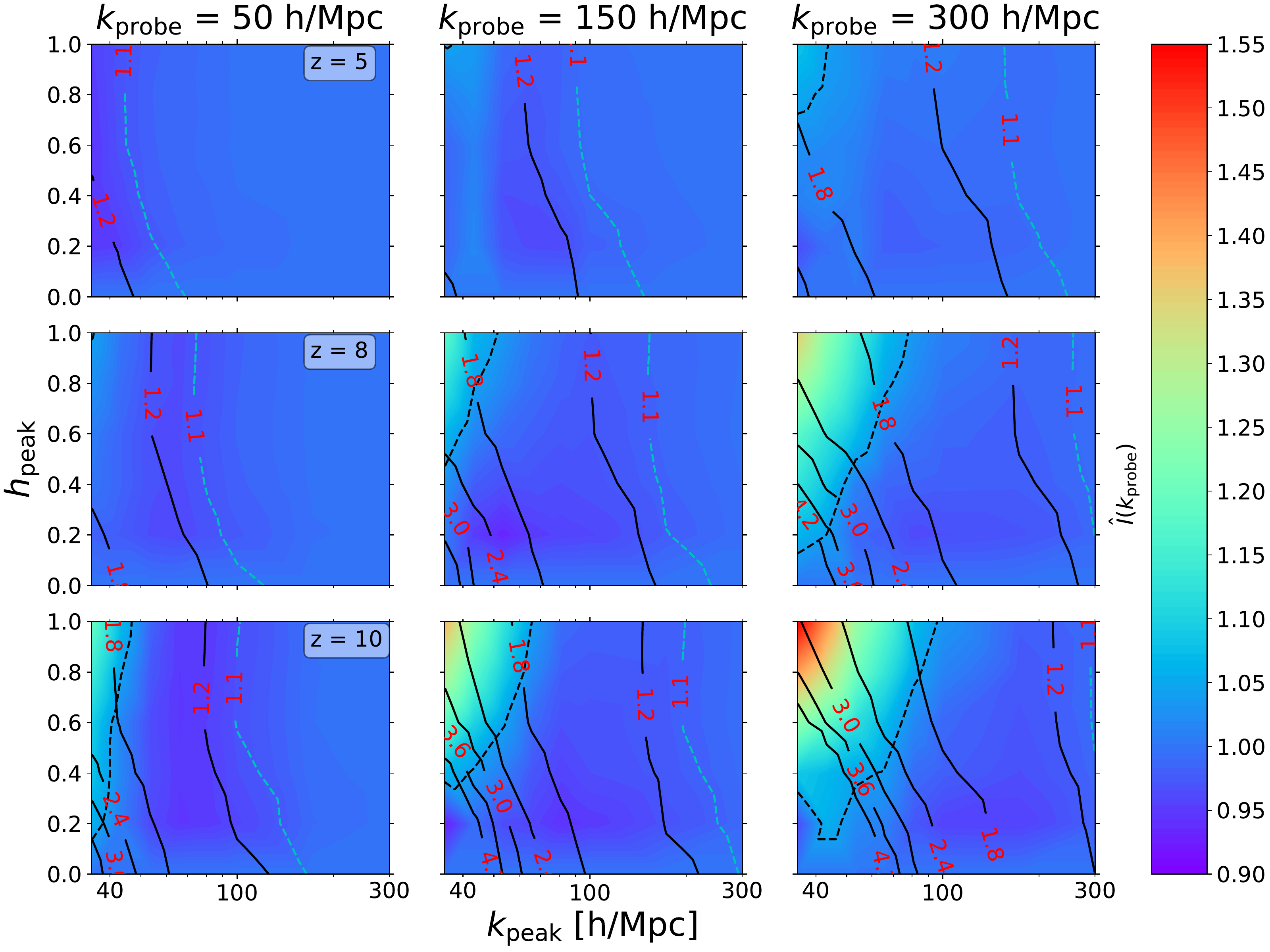}
    \caption{Division of structure formation models in the parameter space $h_{\rm peak}$ and $k_{\rm peak}$ of Eq.~\eqref{eq:parametrization} based on the power spectra at different scales $k_{\rm probe}$ and redshifts. See Fig.~\ref{fig:result} for a description of the different elements in this plot.} 
    \label{fig:result_z}
\end{figure*}

The hashed region on the lower left was constructed taking as a reference the constraints on the thermal WDM particle mass from current Lyman-$\alpha$ forest data from \citet{Murgia2018} using MIKE/HIRES data: $m_{\rm WDM}<3.6$~keV ($2\sigma$ C.L.). To do this, we follow the contour line corresponding to this WDM model, up to the value of $h_{\rm peak}$ that remains degenerate with this WDM model ($h_{\rm peak}\sim0.47$) using a value of $\hat{I}=1.01$ as the dividing threshold. We then continue the Lyman-$\alpha$ constraint line towards larger values of $R$ along this $\hat{I}$ threshold. We remain within this threshold because we expect that beyond, the Lyman-$\alpha$ analysis based on WDM model would no longer be valid due to the impact of the DAO features. In this way, the hashed region on the lower left is our expectation for the exclusion region from Lyman-$\alpha$ data. Drawing this region more precisely would require a full analysis of the predictions of the 1D flux power spectrum within our framework. For the $h=0.2$ ($n=4$) case, this was done in \citet{Archidiacono2019} and we indicate their upper limit of $a_n\xi^4<30{\rm Mpc}^{-1}$ as an upper bound on $k_{\rm peak}$ with the arrow in Fig.~\ref{fig:result}. We notice that their direct constraint on the wDAO model is close to our expectation from models that are degenerate with the WDM constraints (envelope of the hashed region in Fig.~\ref{fig:result}). However, there are a few factors that are likely responsible of the mismatch. Most notably, our estimate is based on the 3D matter power spectrum, while the constraint from \citet{Archidiacono2019} is derived from the 1D flux power spectrum. Finally, we have indicated with a purple star symbol the location of the benchmark ETHOS-4 model defined in \citet{Vogelsberger2016}, which is seemingly barely allowed within the Lyman-$\alpha$ exclusion region. This is a point that was noticed in \citet{Vogelsberger2016} where the linear power spectrum of this model was considered to have a WDM equivalent (in terms of the cutoff) with a thermal particle mass of $m_{\rm WDM}=3.66$~keV. Our results confirm this correspondence and indicate that the ETHOS-4 model has a power spectrum that is nearly indistinguishable at $z=5$ from a WDM model with $m_{\rm WDM}\sim3.5$~keV.

From the slopes of the contour lines in Fig.~\ref{fig:result}, we can see that the degeneracies are not between weak DAO and WDM models with the same power spectrum cutoff (i.e. $k_{\rm peak}$), but that the additional power coming from the DAOs at scales smaller than $k_{\rm peak}$ still matters and can only be accounted for by WDM models with a cutoff at smaller scales. This is shown more clearly on the left panels of Fig.~\ref{fig:degen}, which show the power spectrum for a weak DAO model (ETHOS-2 in \cite{Vogelsberger2016}; $h_{\rm peak}=0.2$,$k_{\rm peak}=46.5$\hmpc) and a WDM model with $m=2.3\,{\rm keV}$ at $z=127$ (top) and $z=5$ (bottom). Despite having distinct linear power spectra, these models are nearly degenerate at $z=5$; they have both the same value of $R(k_{\rm probe}=500$\hmpc) and $\hat{I}(k_{\rm probe}=500$\hmpc). Moreover, in order to match the weak DAO model,
the WDM model needs to sustain more power at larger scales in the linear power spectra to compensate for its steeper cutoff.
The right panels of Fig.~\ref{fig:degen} show the distinct behaviour of a strong DAO model. In this case the additional power from the secondary DAO peaks have an impact in the non-linear power spectrum down to $z=5$ that cannot be replicated by a WDM model: in order for the sDAO and the WDM models to have the same power at $k_{\rm probe}=500$\hmpc, the WDM linear cutoff needs to occur at significantly smaller scales than the one for the sDAO model. Thus, the WDM model has more power at intermediate scales by $z=5$.

Figure~\ref{fig:result_z} is equivalent to Fig.~\ref{fig:result} but at different redshifts
($z=5,8,10$) and scales $k_{\rm probe}$ ($50,150,300$\hmpc). Note that we use the same colour scale for $\hat{I}$ for all cases in order to ease the comparison between the different panels. Focusing on the values of $R(k_{\rm probe})$ represented by the contour lines first, we can see that they shift towards larger $k_{\rm peak}$ as the redshift increases (from top to bottom) or as $k_{\rm probe}$ increases (from left to right). The former trend is expected since at higher redshift the clustering properties have departed less from the linear evolution, where the different DM models differ the most from CDM at all scales. The latter trend is simply due to the damping envelope in the different DM models, which produces an effective cutoff towards smaller scales, and thus $R$ will naturally be larger towards larger values of $k_{\rm probe}$.
Looking at the colour contours, it is apparent that at lower redshifts and/or small $k_{\rm probe}$, none of the DAO models explored are clearly distinguishable from WDM models (i.e. the value of $\hat{I}$ is too close to 1) with the DAO region we highlighted in Fig.~\ref{fig:result} (black dashed line) essentially disappearing in the top left panels. We emphasize that this is independent of the strength of the DAO features in the linear power spectrum. The opposite happens as $k_{\rm probe}$ and/or the redshift increases, the DAO region increases to cover a larger region of the parameter space. This is because at higher redshift the WDM and DAO models become ever more divergent since there is less time to erase the DAO features and equalize the power at all scales. At smaller scales, the initial difference in power was larger and needs more time to get erased, and additionally for large $k_{\rm probe}$, the signal over a wide range of $k$-modes is accumulated.

\begin{figure}
    \centering
    \includegraphics[width=\columnwidth]{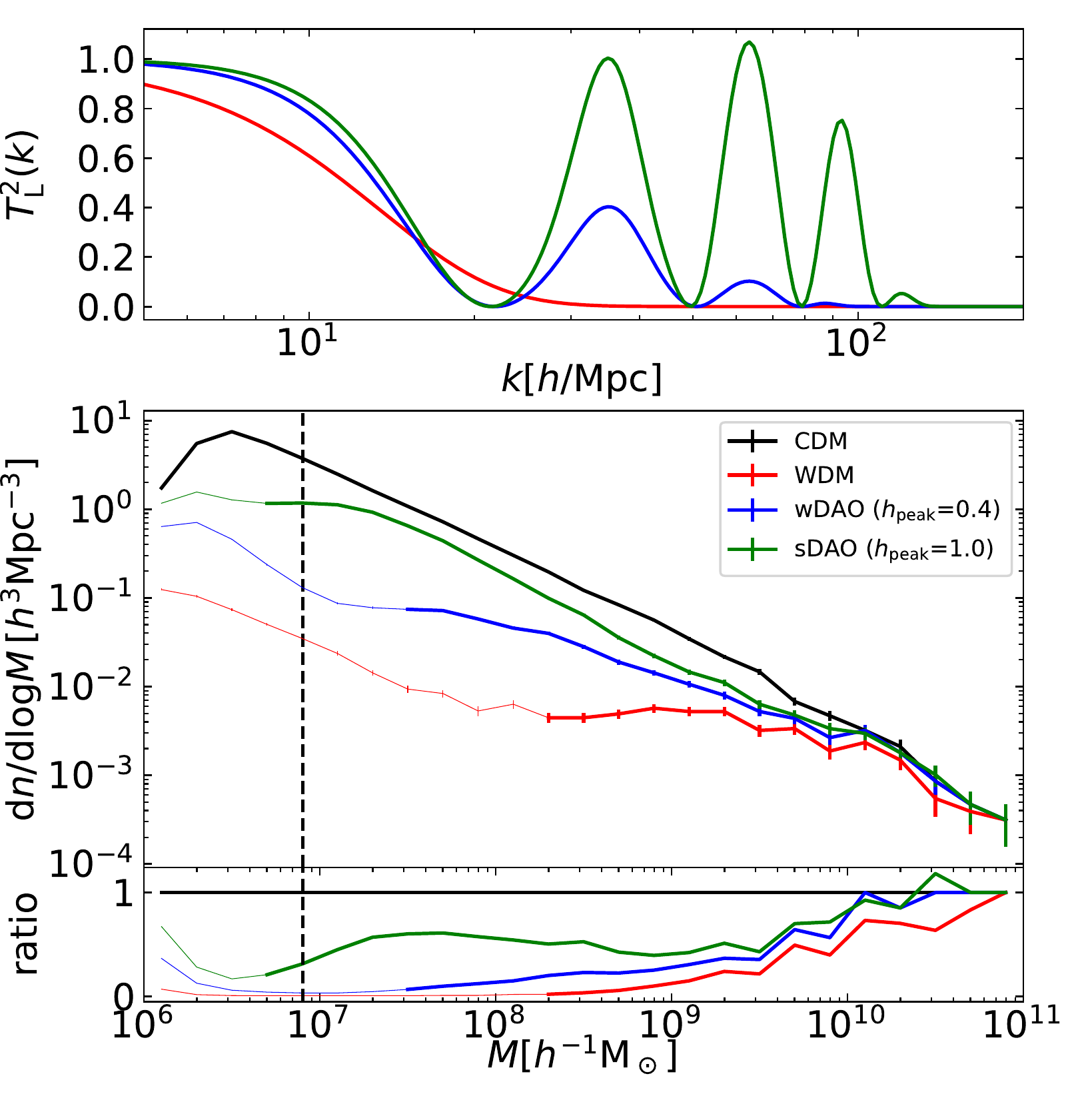}
    \caption[]{{\it Top panel:} Initial transfer function $T_{\rm L}^2(k)$ for examples of the WDM (red), weak DAO (blue) and strong DAO (green) models; all of which have the same value of $k_{\rm peak}=35$\hmpc. {\it Bottom panel:} Halo mass function at $z=5$ for the models above and CDM (black). The error bars denote Poisson counting errors. Masses below the limiting mass (see \citealt{Wang2007}) are indicated by thin lines. To the left of the vertical dashed line haloes have less than 100 simulation particles.}
    \label{fig:mass_fct}
\end{figure}

\subsection{Halo mass function}

The halo mass function provides another relevant measure to characterise structure formation models. It is also more sensitive to the differences across DM models in the linear regime than the non-linear power spectrum since it preserves a stronger memory of the history into collapsed haloes across time \citep[for alternative DM models, including those with DAOs, this was studied e.g. in][]{Leo2018}.

Figure~\ref{fig:mass_fct} shows an example of a comparison of the halo mass function at $z=5$ between the CDM model, and a WDM, weak DAO and strong DAO model with the same cutoff scale in the initial power spectrum. It can be seen that even though the three models are designed to start deviating from CDM at roughly the same mass, the slope of the halo mass function at smaller scales is very different. The halo mass function for the WDM model stays roughly flat towards the left of the cutoff mass until the slope rises again artificially due to the presence of spurious haloes caused by well-known discreteness effects in models with a primordial power spectrum cutoff. The limiting mass below which one can no longer trust the halo mass function is well described by a formula that depends on the cutoff scale of the model and the spatial resolution of the simulation \citep[see][]{Wang2007}. As can be seen in Fig.~\ref{fig:mass_fct}, this formula describes reasonably well the scale at which spurious haloes start to dominate not only for the WDM model, but also for the weak and strong DAO models. For the CDM model, the mass function can be trusted to even lower masses until the simulation particle number is too low to resolve haloes (typically $\lesssim$100). On the other hand, the additional small scale power of the DAO models keeps the slope of the halo mass function steeper at small masses, relative to the WDM case, and for strong DAOs, the halo mass function is even parallel to the CDM case, albeit with a reduced normalization.

\begin{figure*}
    \centering
    \includegraphics[width=\textwidth]{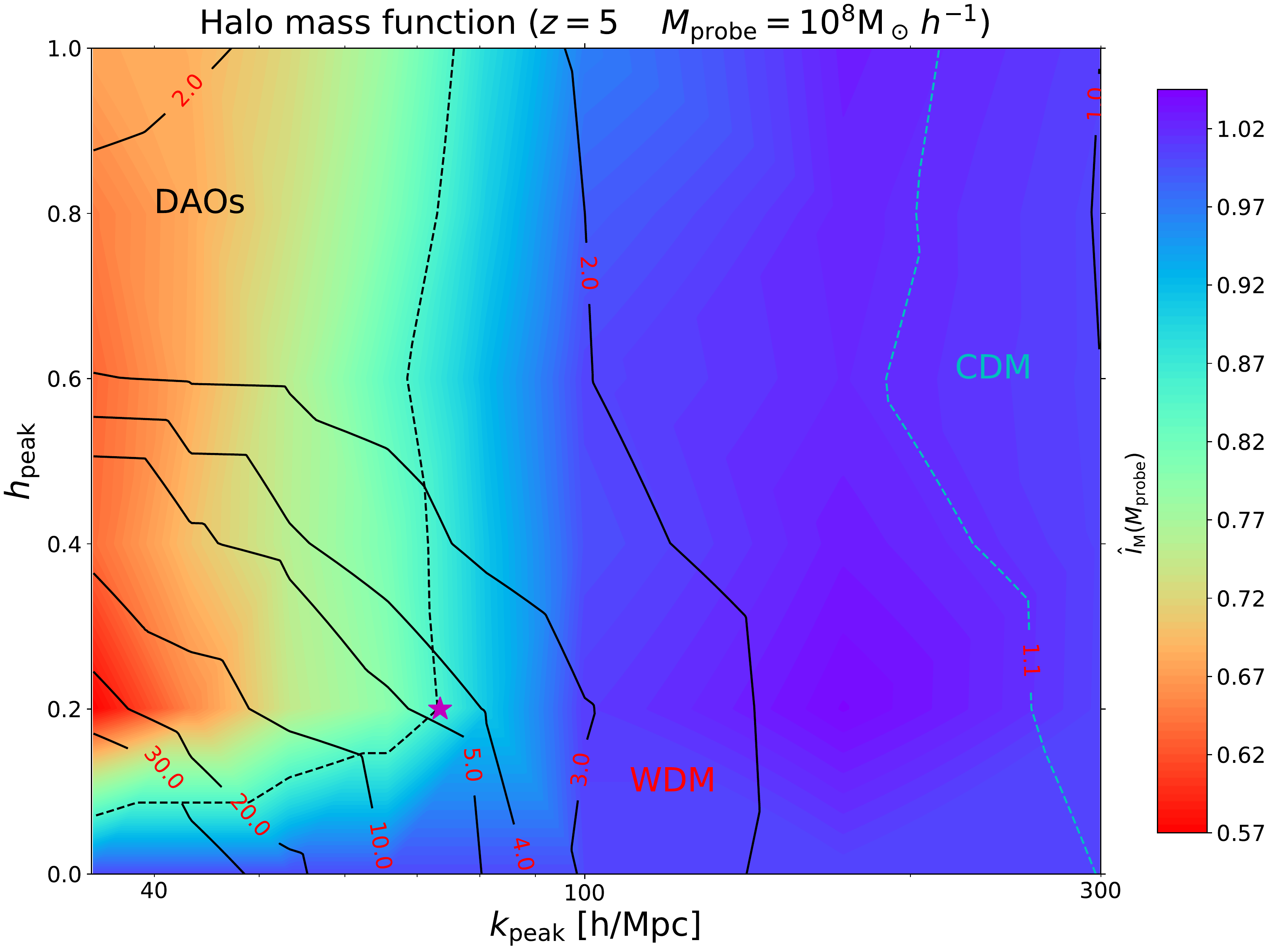}
    \caption[]{Division of structure formation models in the effective parameter space $h_{\rm peak}$ and $k_{\rm peak}$ (see Eq.~\ref{eq:parametrization}) according to their halo mass functions at $z=5$ for $M_{\rm probe}=10^8 {\rm M_\odot}/h$. The contour lines correspond to the ratio $R_{\rm M}(M_{\rm probe})$ of the CDM halo mass function relative to a given model at $M_{\rm probe}$ ($\hat{I}_{\rm M}(M_{\rm probe})$; see Eq.~\ref{eq:ratio_M}). The colour scale shows the number of haloes with $M>M_{\rm probe}$ normalized to a WDM model with the same number of haloes at $M=M_{\rm probe}$ (see Eq.~\ref{eq:I_M}), where a value of 1 corresponds to areas that are degenerate with WDM. The black dashed line ($\hat{I}_{\rm M}(M_{\rm probe})<0.85$) on the left encompasses the area where DAO features survive until $z=5$. The region to the right of the blue dashed line, $R_{\rm M}(M_{\rm probe})=1.1$, can be considered as nearly indistinguishable from CDM down to $M_{\rm probe}$. The ETHOS-4 model used as a benchmark in \cite{Vogelsberger2016} is indicated by the purple star.}
    \label{fig:result_halo}
\end{figure*}

Figure~\ref{fig:result_halo} shows the structure formation models in the parameter space ($h_{\rm peak}$,$k_{\rm peak}$)
as characterised by the halo mass function of our simulations at $z=5$ and at a halo mass of $M_{\rm probe}=10^8 {\rm M_\odot}/h$. This figure is analogous to Fig.~\ref{fig:result}, with the contours showing in this case the ratio between the CDM halo mass function and that of a given model at $M_{\rm probe}=10^8 {\rm M_\odot}/h$: 
\begin{equation}\label{eq:ratio_M}
    R_{\rm M}(M_{\rm probe})=\left(\frac{MF_{\rm CDM}}{MF}\right)_{M_{\rm probe}}
\end{equation}
where $MF=dn/d{\rm log}M$ is the differential halo mass function. On the other hand, the colour contours represent the number of haloes with $M>M_{\rm probe}$ normalized to the number of haloes of a WDM model with the same ratio at $M_{\rm probe}$: 
\begin{equation}\label{eq:I_M}
    \hat{I}_{\rm M}(M_{\rm probe})=\left(\frac{I_{\rm M}(M_{\rm probe})}{I_{{\rm M},{\rm WDM}}(M_{\rm probe})}\right)_{R_{\rm M}(M_{\rm probe})}
\end{equation}
where 
\begin{equation}
    I_{\rm M}(M_{\rm probe})=\displaystyle{\int_{M_{\rm probe}}^{M_{\rm max}}}MF d{\rm log}M
\end{equation}
with $M_{\rm max}=10^{11}{\rm M_\odot}/h$ being the maximum mass for which we can measure the halo mass function.
We can see that the models with $k_{\rm peak}>100$\hmpc are nearly indistinguishable from the corresponding WDM model (that are lying at the same contour line), since in this case the cutoff in the halo mass function is so close to $M_{\rm probe}$
that the different models (irrespective of the value of $h_{\rm peak}$) do not have very different slopes for their halo mass functions yet and thus, they all look alike. 
On the contrary, for $k_{\rm peak}<100$\hmpc, the DAO models have halo mass functions with slopes that are clearly steeper (and thus distinguishable) than that of the WDM model below the cutoff mass. Therefore, in order for the corresponding WDM model to lie on the same contour line (i.e. to have the same halo mass function at $M_{\rm probe}$), it needs to have a cutoff scale at a relatively smaller mass (larger $k_{\rm peak}$), and thus will necessary have more haloes with $M>M_{\rm probe}$ than the DAO model (see Eq.~\ref{eq:I_M}).
We observe that for $k_{\rm peak}\lesssim100$\hmpc, the slope in the halo mass function towards smaller masses is related to $h_{\rm peak}$, which can be seen by looking at how the contour lines bend ever more sharply towards lower values of $k_{\rm peak}$ as $h_{\rm peak}$ increases, eventually becoming nearly flat for
$h_{\rm peak}\sim0.6$ at $k_{\rm peak}\sim 30-60$\hmpc. This implies that for these models, the actual mass cutoff (given by $k_{\rm peak}$) does not matter any longer since they all have the same mass function at $10^8 {\rm M}_\odot h^{-1}$ haloes. Naturally, these models are still distinguishable since they have different halo abundances at larger masses.

In contrast to the division of structure formation models based on the non-linear power spectrum (Fig.~\ref{fig:result}), in this case represented by the halo mass function, the distinctive DAO region (black dashed line in Fig.~\ref{fig:result_halo}) occupies a larger region of the parameter space, reaching into the regime of the weak DAO models. For instance, the red star in Fig.~\ref{fig:result_halo} corresponds to the ETHOS-4 model used in \cite{Vogelsberger2016} and it appears at the border of our definition of the DAO structure formation region. It is thus clear that even though weak DAO models are degenerate with WDM models in their non-linear power spectrum, this degeneracy is broken for the halo mass function.

\begin{figure}
    \centering
    \includegraphics[width=\columnwidth]{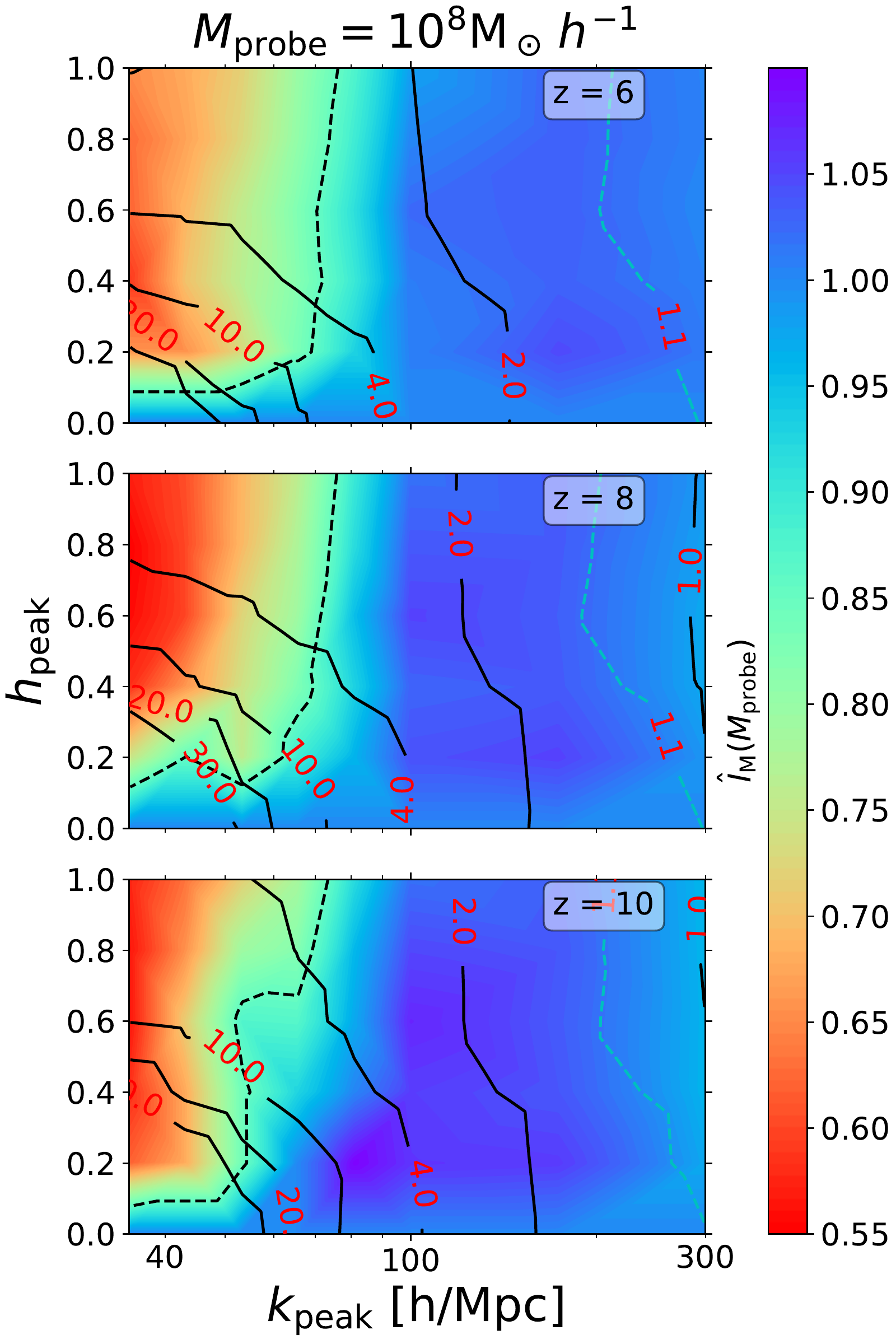}
    \caption{Division of structure formation models in the parameter space $h_{\rm peak}$ and $k_{\rm peak}$ of Eq.~\eqref{eq:parametrization} based on the halo mass functions at $M_{\rm probe}=10^8 {\rm M}_\odot/h$ and different redshifts. See Fig.~\ref{fig:result_halo} for a description of the different elements in this plot.} 
    \label{fig:result_halo_z}
\end{figure}

At different redshifts (Fig.~\ref{fig:result_halo_z}), the DAO region (black dashed line) remains almost unchanged (shrinking slightly at high redshift); the same is true for the CDM-like region (blue dashed line). It is only the contour lines of constant $R_{\rm M}$ that change across redshift, with the ratio of the halo mass function at $10^8 {\rm M}_\odot/h$ becoming larger at higher redshift for all the region below $k_{\rm peak}\lesssim100$\hmpc.
Therefore, the halo mass function provides a diagnostic to classify structure formation models that is less susceptible to being erased by the non-linear evolution than the power spectrum. A more detailed analysis of the halo mass function for the DAO models studied here will be presented in the future.

\section{Conclusions}
\label{sec:conclusion}

There are multiple ways in which non-standard DM physics can introduce a cutoff in the linear matter power spectrum whose shape can range from an exponential featureless free-streaming collisionless damping (as in thermal WDM models) to a shallower collisional damping driven by DM-dark radiation interactions with strong DAOs. The effective theory of structure formation (ETHOS), introduced in \cite{Cyr-Racine2016} aims at connecting the particle physics parameters of a variety of DM models into effective parameters that characterise the linear power spectrum. In this way, DM particle models can be classified in terms of a set of parameters that fully describe the linear power spectrum, particularly the characteristics of the small-scale cutoff and DAOs. It is however, not trivial to characterise the signature that these different departures from the linear CDM power spectrum leave in the non-linear regime of structure formation. It may indeed be possible that the gravitational coupling between different scales erases features like the DAOs making all models essentially indistinguishable from the standard WDM cutoff at the scales that are relevant for galaxy formation and evolution\footnote{Relevant departures from CDM are bounded at large scales by current constraints based on e.g. observations of the Lyman-$\alpha$ 1D flux spectrum, and at small scales by the suppression of galaxy formation below the atomic cooling limit.} ($10$\hmpc$\lesssim k\lesssim500$\hmpc). In this work 
we address this question by performing a 
large number of cosmological simulations within the ETHOS framework. Our goal is to define a reduced set of simple yet physically motivated parameters that allow to distinguish DM models based on differences in how structure formation proceeds (at the scales relevant for the physics of galaxies). In this first work, we have concentrated on DM-only simulations at high redshift ($z\geq5$).
Our analysis and main results can be summarised as follows:

\begin{itemize}
    \item We have implemented a zoom-in simulation technique to efficiently cover a wide range of scales ($0.2-500$\hmpc) and accurately reconstruct the (average) matter power spectrum in this range (see Fig.~\ref{fig:performance}). The computational cost of this method is significantly less than a uniform simulation with equivalent range (see Table~\ref{tab:resources}).
    \item We introduced a new analytic formula (Eq.~\eqref{eq:parametrization}, Fig.~\ref{fig:parametrization}) to describe the linear transfer function (relative to CDM) of models with a primordial cutoff, which accommodates both WDM and models with DAOs. This formula is accurate enough to reproduce the non-linear power spectrum for the scales of interest compared to the full calculation with a Boltzmann code 
    (see Fig.~\ref{fig:param1}). Crucially, only two free parameters in this formula, $\{h_{\rm peak},k_{\rm peak}\}$ the amplitude and scale of the first DAO peak, are sufficient to characterise non-linear structure formation at high-redshift for WDM\footnote{In the case of WDM, $k_{\rm peak}$ is connected to $k_{1/2}$, the scale at which the transfer function squared is equal to $1/2$; see Eq.~\eqref{eq:alpha}.} and a large class of ETHOS models with DAOs. Moreover, we
    found a simple physical interpretation for these two main parameters where $k_{\rm peak}$ is connected to $\eta_\chi$ the time of DM decoupling from the DR (see Fig.~\ref{fig:kpeak}) and $h_{\rm peak}$ is determined by the ratio of the DM drag opacity to the Hubble rate at the time of DM decoupling $\dot\kappa(\eta_\chi)/\mathcal{H}$ (see Fig.~\ref{fig:h}).
    \item Using 50 simulations down to $z=5$ within the \{$h_{\rm peak},k_{\rm peak}$\} parameter space (a new effective space in ETHOS), we have been able to classify DM models into regions with distinct non-linear structure formation at galactic scales (CDM, WDM, DAOs), quantified by the non-linear power spectrum and the halo mass function at high redshift. 
    \item As far as the non-linear matter power spectrum is concerned, we find that only a small region within this effective parameter space, corresponding to relatively small values of $k_{\rm peak}$ and large values of $h_{\rm peak}$ (strong DAO models), still preserves a signature of the DAOs at $z=5$ at the galactic scales corresponding to the smallest galaxy-forming haloes $~500$\hmpc (Fig.~\ref{fig:result}). The rest of the relevant parameter space including weak and strong DAOs is either degenerate with WDM, which we quantify with $\hat{I}(k_{\rm probe})$ (Eq.~\ref{eq:iprobe2}), or indistinguishable from CDM. This distinct DAO region expands at higher redshifts and contracts at smaller scales (see Fig.~\ref{fig:result_z}).
    \item We find that it is possible to break (to a certain extent) the degeneracies between weak DAO models (small values of $h_{\rm peak}$) and WDM models seen in the non-linear power spectrum by characterising structure formation models using the halo mass function instead. This is because the halo mass function retains a memory of the linear power spectrum, having a slope that is very sensitive to the value of $h_{\rm peak}$. In this way, the distinct DAO region covers a much larger region of the parameter space and changes only slightly with redshift (see Figs.~\ref{fig:result_halo}$-$\ref{fig:result_halo_z}).
\end{itemize}

Using our results, it is possible to use the new analytic formula we propose (Eq.~\ref{eq:parametrization}) to fit the linear power spectrum of a broad class of ETHOS models with DAOs, and use the values of \{$h_{\rm peak},k_{\rm peak}$\} to determine to which structure formation region they belong to in the non-linear high-redshift regime (CDM-like, WDM-like or DAO), without performing additional $N$-body simulations. Notice that this is valid for any DM particle model with a primordial power spectrum with DAOs that can be fitted accurately with our formula (up to the second DAO peak). In other words, given the values of \{$h_{\rm peak},k_{\rm peak}$\}, our method allows to infer the value of the non-linear power spectrum and halo mass function at any relevant scale/mass at high-redshift $z\sim10-5$. 

Furthermore, The effective parameters for structure formation we propose here represent a potentially powerful way to constrain the parameter space of a variety of particle physics models by using observations in the high-redshift Universe, such as the Lyman-$\alpha$ 1D flux power spectrum. We should remark however that in order to accurately exploit this avenue, we need to incorporate the baryonic physics that is relevant for the intergalactic medium into the ETHOS parameter space, which is something we plan to do in the future. However, the results for the $h_{\rm peak}=0.2$ case presented in \citet{Archidiacono2019} and the expectation based on WDM constraints \citep{Murgia2018} indicate that the lower left region in Fig.~\ref{fig:result} is likely in tension with current Lyman-$\alpha$ data. Another avenue we will explore is to extend our results towards lower redshifts, where DM self-interactions have a significant impact in the centre of DM haloes, and thus need to be incorporated as an additional parameter to classify structure formation regimes.

% The last numbered section should briefly summarise what has been done, and describe
% the final conclusions which the authors draw from their work.

\section*{Acknowledgements}

SB and JZ acknowledge support by a Grant of Excellence from the Icelandic Research Fund (grant number 173929). CP acknowledges support by the European Research Council under ERC-CoG grant CRAGSMAN-646955. The simulations were performed on resources provided by the Icelandic High Performance Computing Centre at the University of Iceland, and the Odyssey cluster supported by the FAS Division of Science, Research Computing Group at Harvard University.

\section*{Data availability}
The data underlying this article will be shared on reasonable request to the corresponding author.

%%%%%%%%%%%%%%%%%%%%%%%%%%%%%%%%%%%%%%%%%%%%%%%%%%



\bibliographystyle{mnras}
\bibliography{refs} % if your bibtex file is called example.bib

%\begin{thebibliography}{99}
%\end{thebibliography}

%%%%%%%%%%%%%%%%%%%%%%%%%%%%%%%%%%%%%%%%%%%%%%%%%%

%%%%%%%%%%%%%%%%% APPENDICES %%%%%%%%%%%%%%%%%%%%%

\appendix

\section{Convergence tests}
\label{sec:convergence}

\begin{figure}
    \centering
    \includegraphics[width=\columnwidth]{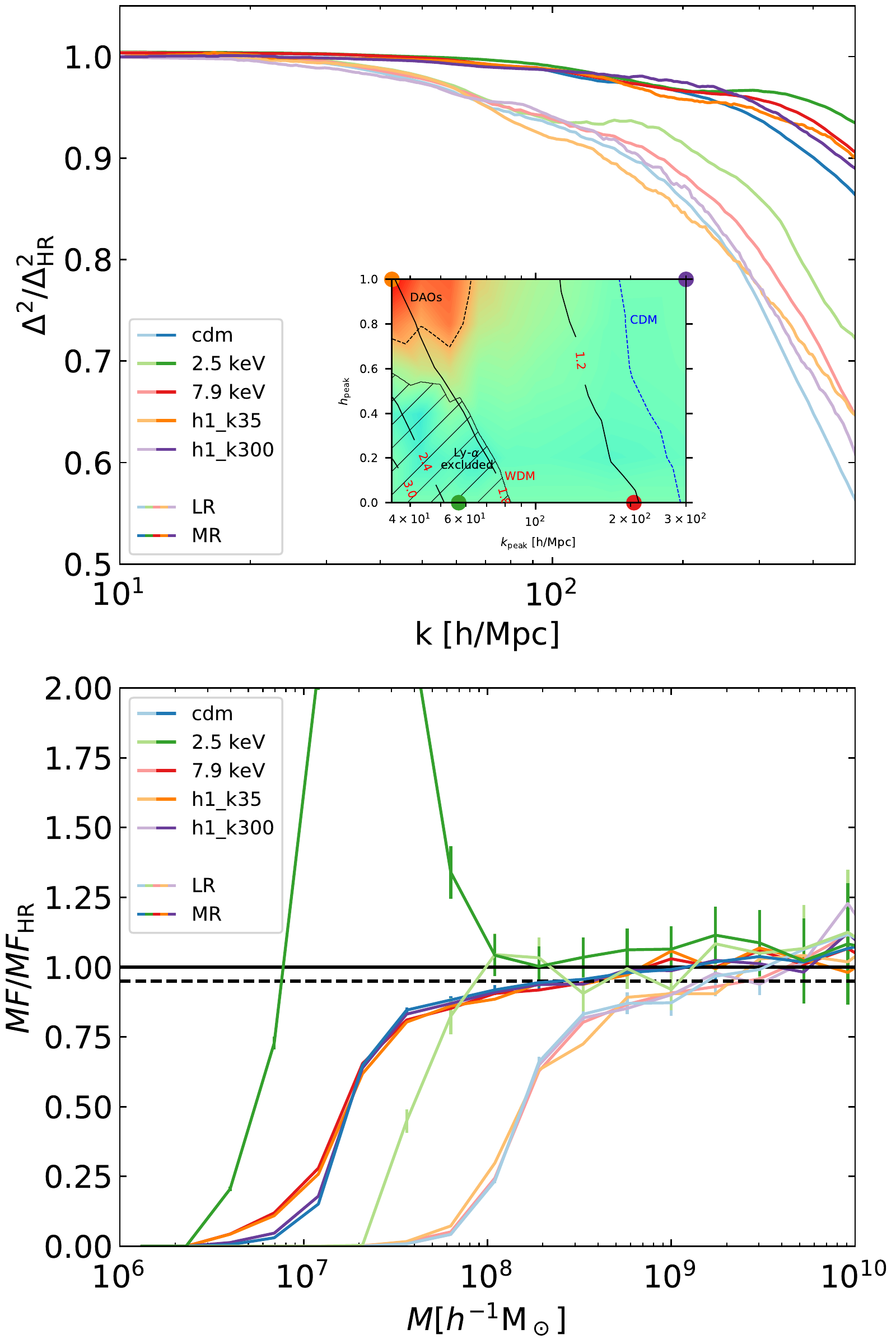}
    \caption{Top panel: Convergence of the power spectrum for 5 models: CDM (blue) and four models marked with solid circles in inset, which is a reproduction of the parameter space plot Fig.~\ref{fig:result}. The vertical axis is the ratio of the dimensionless power spectrum of the low-resolution (faded lines) and medium resolution (solid lines), relative to that of the highest resolution run at $z=5$. The horizontal shaded line marks a convergence level of $5\%$. Bottom panel: The same as the top panel but for the halo mass function.}
    \label{fig:convergence}
\end{figure}

To determine the minimum scale at which we can trust our measurements of the power spectrum and the halo mass function in our simulations, we performed convergence tests for a few DM models using three resolution levels for each. These models cover representative regions of the parameter space we explore and, based on our analysis, they bracket the possible range of convergence variations. The convergence reported here is thus a fair representation of the convergence for all the parameter space explored in this paper.
The three resolution levels were done within a $\sim$($6.25$\mpch)$^3$ Lagrangian zoom region at $z=5$ with $1024^3$ (LR), $2048^3$ (MR), and $4096^3$ (HR) effective particle resolution. We set the goal to determine for the two lower resolution levels, the wavenumber $k_{\rm conv}$ at which the power spectrum differs by $5\%$ with respect to the highest resolution. The upper panel of Fig.~\ref{fig:convergence} shows this convergence test for the models highlighted with large circles within the parameter space shown in the inset. Notice that since

we subtract the shot-noise from the power spectra, the power falls off at the smallest scales. This behaviour is responsible for a smaller convergence scale (i.e. larger $k_{\rm conv}$) for the WDM models (green and red lines) compared to the sDAO models; this also applies in general to all wDAO models. 
In models with a steep {\it linear} power spectrum cutoff, the non-linear true power is expected to be highly suppressed at sufficiently small scales and thus, there is not much power left at the unresolved scales.  
However, the sDAO case is similar to CDM, there is still significant power left at the unresolved scales and therefore, the $5\%$ convergence level is reached at larger scales (smaller $k_{\rm conv}$). Despite this difference across different DM models, Fig.~\ref{fig:convergence} shows that there is at least a factor of $3$ improvement in $k_{\rm conv}$ between the LR and MR simulations. The power spectra of the latter are converged to $k_{\rm conv}> 250$\hmpc and thus, assuming at least another factor of 2 improvement for the HR simulations, all models are converged to better than $5\%$ at $k\sim500$\hmpc. 

The bottom panel of Fig.~\ref{fig:convergence} is equivalent to the top panel but for the halo mass function. It shows the ratio of the halo mass functions of the two lower resolution levels to that of of the high resolution. It can be seen that the LR simulations drop below $5\%$ convergence at $\sim 2\times10^9 {\rm M}_\odot/h$, while for the MR simulation this threshold occurs at $\sim 3\times10^8 {\rm M}_\odot/h$. We highlight however, that for the $2.5\,{\rm keV}$ model discreteness effects cause the well-known effect of spurious haloes \citep{Wang2007}, which appear in this case at $M<10^8 {\rm M}_\odot/h$, dominating the signal. Although all models with a primordial power spectrum cutoff suffer from spurious haloes we find that in all cases, for our highest resolution, we are free from this effect at a halo mass of $10^8 {\rm M}_\odot/h$. We therefore set this mass as our lower mass limit for all cases and report a convergence of the halo mass function to better $5\%$ for this and larger masses. While the resolution based convergence discussed above affects the small mass end of the halo mass function, the high mass end is affected by the limited volume of the zoom-in region, 
which can only encompass a few of the most massive haloes, leading to large Poisson (counting) errors. However, this is not relevant for our purposes as our models converge at large masses anyway and the differences we are interested in appear at smaller halo masses.

\section{The DM linear power spectrum in the tight DM-DR coupling limit}
\label{sec:linpk}

The goal in Section~\ref{sec:connection} is to connect two different parametrizations for the linear power spectrum. On the one hand, the parameters we have defined in this work $\{h_{\rm peak}, k_{\rm peak},h_2\}$, and on the other the parameters used in \citet{Cyr-Racine2016}, essentially $\{a_n, n,\alpha_{l\geq2}\}$. As we show in Section~\ref{sec:results}, the former set can be used directly to quantify the differences between  different structure formation models in the non-linear regime, while the latter can be connected directly to the particle physics parameters of a given model.
In this Appendix we explore the connection between these two sets of parameters, which allows to obtain a physical interpretation of the final parametrization we have used in this work.

We start by recalling that in \cite{Cyr-Racine2016} the DR and DM-drag opacities for a given ETHOS model are respectively written as:
\begin{equation} \label{eq:power}
    \begin{aligned}
        \dot\kappa_{\rm DR}(z) &= -(\Omega_\chi h^2) a_n \left(\frac{1+z}{1+z_{\rm D}}\right)^n \\
        \dot\kappa_\chi(z) &= -\frac{4}{3} (\Omega_{\rm DR} h^2) a_n \frac{(1+z)^{n+1}}{(1+z_{\rm D})^n},
    \end{aligned}
\end{equation}
if we just consider a single pure power law, where $\Omega_\chi$ and $\Omega_{\rm DR}$ are the DM and DR densities in units of the critical density, $h$ is the dimensionless Hubble constant, $a_n$ and $n$ are given by the particle physics model, and $z_{\rm D}$ is an arbitrary normalization factor\footnote{This is the numerical value that was chosen in \cite{Cyr-Racine2016} to be the redshift when the DM opacity becomes equal to the conformal Hubble rate.} chosen to be $z_{\rm D}=10^7$. Eqs.~\eqref{eq:power} is an approximation that assumes that a single power law is sufficient for a given value of $n$, which is a good approximation for instance in the case of DM interacting with massless radiation via a massive mediator (models of this type were simulated in \citealt{Vogelsberger2016}). Assuming that DM and DR decouple well within the radiation dominated era (where the Hubble rate, relative to the conformal time goes as $\mathcal{H}=\eta^{-1}$; note that we use the conformal time in units of {\rm Mpc}), Eqs.~\eqref{eq:power} can be rewritten in terms of the conformal time, using $1+z\propto\eta^{-1}$,
\begin{equation}\label{eq:B2}
    \begin{aligned}
        \dot\kappa_{\rm DR}(\eta) &= -(n-1) \frac{\eta_{\rm DR}^{n-1}}{\eta^n} \\
        \dot\kappa_\chi(\eta) &= \frac{\dot\kappa_{\rm DR}}{\mathcal{R}} = -(n-1)\eta_1 \frac{\eta_{\rm DR}^{n-1}}{\eta^{n+1}} ,
    \end{aligned}
\end{equation}
where $\eta_{\rm DR}$ is the conformal time of DR decoupling defined by $\int_{\eta_{\rm DR}}^{\eta_0} -\dot\kappa_{\rm DR} d\eta=1$, and $\eta_1$ is the conformal time when $\mathcal{R} = \frac{3}{4}\rho_{\rm DM}/\rho_{\rm DR} = \eta/\eta_1=1$. Note that this definition of the decoupling time takes the weakly coupled regime into account, i.e., a broad visibility function will lead to a later decoupling time. The DR decoupling time $\eta_{\rm DR}$ can be related directly with $a_n$ and $n$:
\begin{equation}\label{eq:eta_DR}
    \eta_{\rm DR} = \left[\frac{\Omega_\chi h^2 a_n}{n-1} \left(\frac{1}{H_0 \Omega_{\rm r}^{1/2} (1+z_{\rm D})}\right)^n \right]^\frac{1}{n-1} ,
\end{equation}
where $\Omega_{\rm r}$ is the radiation density in units of the critical density, and $H_0$ is the Hubble constant today. In a similar way, we can define the conformal time for DM decoupling by $\int_{\eta_\chi}^{\eta_0} -\dot\kappa_\chi d\eta=1$, and thus $\eta_\chi$ can be written as: 
\begin{equation}\label{eq:eta_chi}
    \eta_\chi = \left(\frac{n-1}{n}\eta_1 \eta_{\rm DR}^{n-1}\right)^\frac{1}{n}.
\end{equation}

We can use the tight coupling approximation (e.g. \citealt{Hu1996}) for the DM-DR plasma to roughly capture the acoustic oscillations in the DM fluid as well as the effect of DR diffusion damping. We then propose that the DM density fluctuations have the following $k-$dependence: 
\begin{equation}
    \delta_\chi(k) \propto {\rm cos}(kr_{\rm DAO}) e^{-k^2/k_{\rm D}^2} , \label{eq:delta_chi}
\end{equation}
where the sound horizon scale is given by:
\begin{equation}
    r_{\rm DAO} = \int_0^{\eta_\chi} c_{\rm s} d\eta \approx \frac{\eta_\chi}{\sqrt{3}} ,
\end{equation}
where $c_{\rm s}$ is the dimensionless DM sound speed. 
The parameter $k_{\rm D}$ in Eq.~\eqref{eq:delta_chi}, which controls the damping scale, is given by the tight coupling dispersion relation \citep{Dodelson2003}:
\begin{equation}
    \frac{1}{k_{\rm D}^2} = \int_0^{\eta} \frac{d\eta'}{6(1+\mathcal{R})\dot\kappa_{\rm DR}}\left[\frac{4}{5\alpha_2}+\frac{\mathcal{R}^2}{1+\mathcal{R}}\right] ,
\end{equation}
where $\alpha_2$ is the ratio between the opacity of the quadrupole and dipole moment of the DR multipole hierarchy given by the angular dependence of DM-DR scattering.

For the limit when the visibility function is given as a delta function, the exponential term in Eq.~\eqref{eq:delta_chi} gives the damping envelope of the DAOs, but due to a finite width of the visibility function, the damping envelope should be weighted by the DR visibility function $g_{\rm DR}=-\dot\kappa_{\rm DR} \exp{(-\kappa_{\rm DR})}$: 
\begin{equation}\label{eq:difussion}
    D_n(k) = \int_0^{\eta_0} d\eta g_{\rm DR}(\eta)e^{-k^2/k_{\rm D}^2} ,
\end{equation}
Therefore, under these approximations the linear transfer function is given by:
\begin{equation} \label{eq:analytic_approach}
    %\Delta^2_{\rm ETHOS}(k) 
    T^2_{\rm L}(k)\approx \cos^2{\left(\frac{k\eta_\chi}{\sqrt{3}}\right)} D_n^2(k), %\Delta^2_{\rm CDM}(k) , 
\end{equation}
where the input parameters ($\eta_\chi, \eta_{\rm DR}, k_{\rm D}$) are ultimately given by the cosmological parameters assumed ($\Omega_\chi$, $\Omega_{\rm r}$, $H_0$) and the set of values $\{\Omega_{\rm DR},n,a_n,\alpha_2\}$ for a given ETHOS model.

\begin{figure}
\centering
\includegraphics[width=\columnwidth]{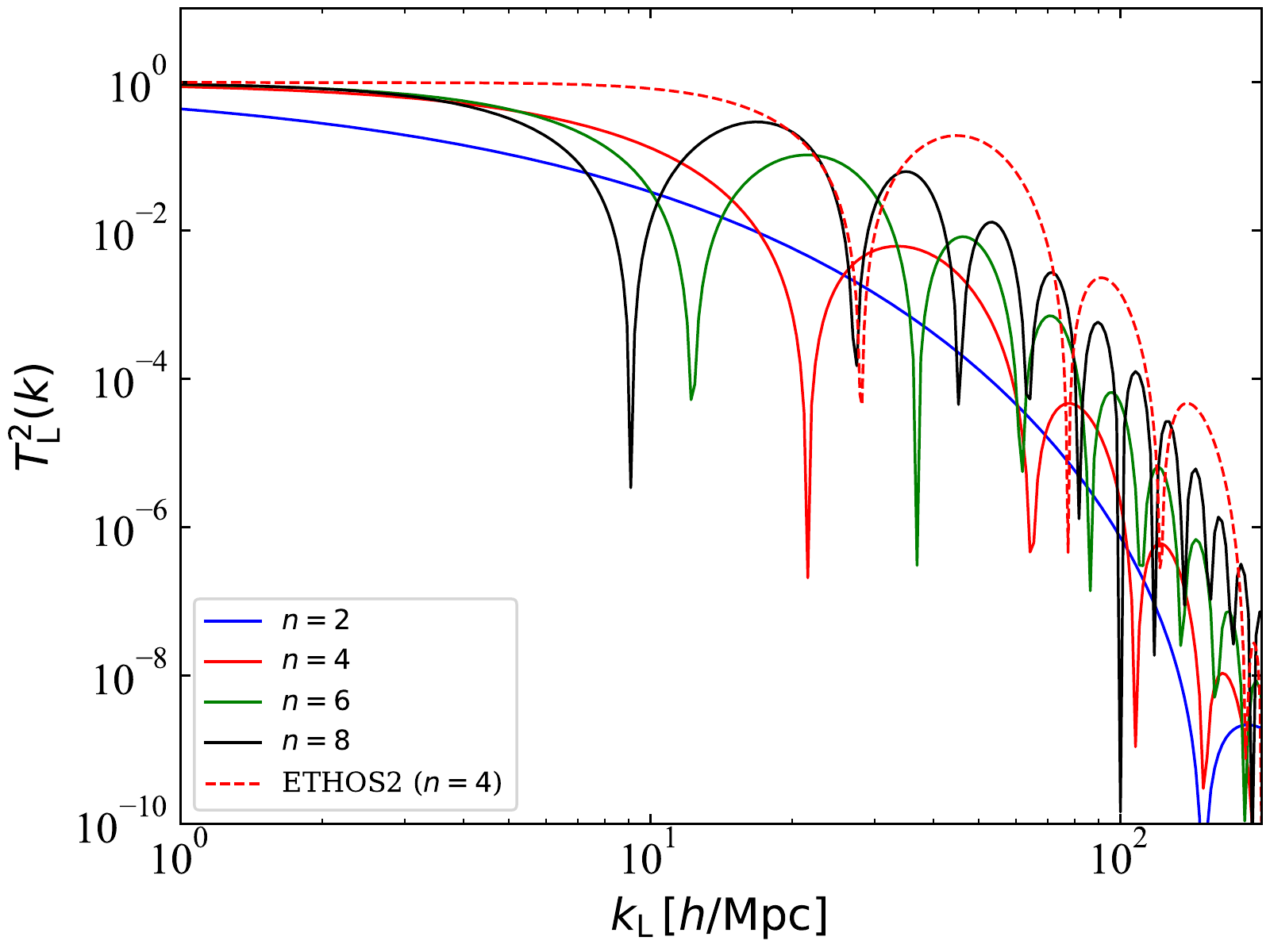}
\caption[]{The solid lines are the linear transfer function for different values of $n$ computed with Eq.~\eqref{eq:analytic_approach} using the a fixed value of $\eta_{\rm DR}$ chosen to match that of the ETHOS-2 model in \cite{Vogelsberger2016} (dashed lines) and ETHOS-2. Note that by fixing $\eta_{\rm DR}$, the value of $a_n$ for a given $n$ is given by Eq.~\eqref{eq:eta_DR}.}
\label{fig:transfer}
\end{figure}

Eq.~\eqref{eq:analytic_approach} results in damped DAOs that resemble the behaviour of the linear power spectrum of ETHOS models generated by the Boltzmann solver (see Fig.~\ref{fig:transfer}). This approach also predicts correctly that the amplitude of the DAOs scales with the value $n$. Unfortunately, the agreement is only qualitative, neither the position of the peaks nor the damping envelope agree with the full calculation. An example of this can be seen by comparing the red solid and dashed lines in Fig.~\ref{fig:transfer}, the latter of which is the ETHOS-2 model in \cite{Vogelsberger2016} (with $n=4$, $a_4=1784.05$\hmpc, $\alpha_{l\geq2}=3/2$; see their table 1).

The previous approach ignores the effect of the velocity perturbations on the density perturbation $\delta_\chi$, the so-called velocity overshoot.
We improved the modelling by including this effect which is part of a more rigorous treatment (see \citealt{Hu1996}). 
Modelling the velocity perturbation as $\theta_\chi\sim {\rm sin}(kr_{\rm DAO})D_n(k)$ moves the first DAO peak to better agree with the numerical results, but the other peaks and damping envelope were still not in agreement. Exchanging the sine and cosine functions for the full analytic solutions, in which case the potentials are given by Bessel functions, gives the evolution of $\delta_\chi$ and $\theta_\chi$ until DR decoupling with good accuracy. The position of the peaks of the DAOs are captured quite accurately with this modification, but the approach still does not capture the transition from the tightly coupled to the weakly coupled regimes correctly. The main reason for this seems to be that the exponential diffusion damping in Eq.~\eqref{eq:difussion} is not an accurate representation of the numerical results for the first few DAO peaks at large scales; it is only a good approximation at much smaller scales. The damping of the first DAO peaks deviates strongly from the exponential behaviour because the timescale for DM decoupling is large compared to the oscillation frequency causing the DM to spend a longer time in the weakly coupled regime.
Trying to model the DM decoupling timescale by weighting $\delta_\chi$ and $\theta_\chi$ with the visibility function, improves slightly the result of this analytical approach, but it remains inaccurate. 

Instead of increasing the complexity of the modelling, which would eventually take us closer and closer to a full approach of the Boltzmann solver, but would defeat the purpose of having a simple physical interpretation, we choose instead to use a phenomenological approach as described in Section~\ref{sec:connection}. The starting point is to notice that the ratio $\dot{\kappa}_\chi/\mathcal{H}\propto(1+z)^n$ is the relevant ratio of timescales (or length scales) in the DM decoupling process, with $\dot{\kappa}_\chi/\mathcal{H}\gg1$ corresponding to the tightly coupled regime and $\dot{\kappa}_\chi/\mathcal{H}\ll1$ to the decoupled regime. We found that the value of this ratio at the DM decoupling time $\eta_\chi$ is actually strongly correlated with the amplitude of the first DAO peak $h_{\rm peak}$ (Fig.~\ref{fig:h}) since it is actually equal to $n$ as can be seen through Eqs.~(\ref{eq:B2}-\ref{eq:eta_chi}), which controls how fast the transition is from the tightly coupled to decoupled regimes, and thus how narrow the DM drag visibility function is. 

% If you want to present additional material which would interrupt the flow of the main paper,
% it can be placed in an Appendix which appears after the list of references.

%%%%%%%%%%%%%%%%%%%%%%%%%%%%%%%%%%%%%%%%%%%%%%%%%%

% Don't change these lines
\bsp	% typesetting comment
\label{lastpage}
\end{document}